\documentclass[reqno]{amsart}
\usepackage{amsthm,amssymb,amsfonts}
\usepackage{bm}

\usepackage{hyperref}
\usepackage[citation-order]{amsrefs}

\usepackage{pstricks,pst-node,pst-plot}
\usepackage{graphicx}

\numberwithin{equation}{section}

\psset{unit=3ex}

\newpsobject{ed}{pcline}{}
\newpsobject{th}{psline}{linewidth=.15}

\newcommand{\s}{\sigma}
\newcommand{\w}{\omega}
\newcommand{\lam}{\lambda}
\newcommand{\de}{\partial}
\newcommand{\cc}{\mathcal{C}}
\newcommand{\eee}{\mathcal{E}}
\newcommand{\denne}[2]{\frac{{\rm d}^#2}{{\rm d} #1 ^#2}}

\def\smfrac#1#2{{\textstyle\frac{#1}{#2}}}

\newcommand{\be}{\begin{equation}}
\newcommand{\ee}{\end{equation}}

\numberwithin{equation}{section}

\newcommand{\rme}{\mathrm{e}}

\newcommand{\rmd}{\mathrm{d}}

\newtheorem{theorem}{Theorem}[section]

\newtheorem{conj}[theorem]{Conjecture}
\newtheorem{assum}[theorem]{Assumption}

\newcommand{\fraka}{\mathsf{a}}
\newcommand{\frakb}{\mathsf{b}}
\newcommand{\frakc}{\mathsf{c}}

\begin{document}

\title[Arctic curves of the six-vertex model\ldots]
{Arctic curves of the six-vertex model\\
on generic domains: the Tangent Method}

\author{F.~Colomo}
\address{INFN, Sezione di Firenze\\
Via G.~Sansone 1, 50019 Sesto Fiorentino (FI), Italy}
\email{colomo@fi.infn.it}

\author{A.~Sportiello}
\address{LIPN, and CNRS, Universit\'e Paris 13, Sorbonne Paris Cit\'e,
99 Av.~J.-B.~Cl\'ement, 93430 Villetaneuse, France}
\email{andrea.sportiello@lipn.univ-paris13.fr}

\begin{abstract}
  We revisit the problem of determining the Arctic curve in the
  six-vertex model with domain wall boundary conditions.  We describe
  an alternative method, by which we recover the previously
  conjectured analytic expression in the square domain.  We adapt the
  method to work for a large class of domains, and for other models
  exhibiting limit shape phenomena.  We study in detail some examples,
  and derive, in particular, the Arctic curve of the six-vertex model
  in a triangoloid domain at the ice-point.
\end{abstract}

\maketitle
\section{Introduction}
\label{sec.intro}

\noindent
Statistical mechanics models in two dimensions with a discrete
symmetry group, within a pure phase, usually show a
spatially-homogeneous order parameter and independence from the
boundary conditions \cite{DG-72}. This can be understood by simple
entropic arguments on local excitations.  The prototype example is the
Ising Model, where the broken symmetry group is just $\mathbb{Z}_2$.

Nonetheless, certain models, characterised by the presence of
conservation laws, under particular conditions may break this paradigm
and show \emph{phase-separation phenomena} and the emergence of a
limit shape \cite{KO-05,RP-10}. In this case we may have spatial
dependence of the order parameter, a strong dependence from the
boundary conditions, and even \emph{frozen regions}, in which the
local entropy vanishes. This is now possible because the conservation
law forbids local excitations on frozen-region vacua, the smallest
perturbations taking the form of a directed path which, in each
direction, shall either reach the boundary, or a non-frozen
(\emph{liquid}) region.  The interface between frozen and liquid
regions, for a given model in a given domain, is called \emph{Arctic
  curve}. The challenge of its determination is the subject of the
present paper.

Among the models presenting phase separation and limit shape
phenomena, those amenable to discrete free fermions are the most
widely studied.  Early examples include Young diagrams with the
Plancherel measure \cite{KV-77}, the evaporation of a cubic crystal
\cite{NienhuisCrystal,CK-01, FS-03}, domino tilings of the Aztec
diamond, \cite{JPS-98}, boxed plane partitions \cite{CLP-98}, Schur
processes \cite{OR-01}. These examples may all be viewed as dimer
models on planar bipartite graphs, for which a general theory has been
constructed \cite{KO-06,KO-05,KOS-06}; other approaches exist for
certain subclasses of models \cite{CJ-14,Pet-14,BK-16,BBCCR-15}.  An
interesting connection between limit shape phenomena in such models
and the out-of-equilibrium evolution of one-dimensional quantum
free-fermion models has been recently unveiled \cite{ADSV-15}.  

Other free-fermionic models presenting a similar phenomenology are
defined in terms of iterated transformations applied to a
deterministic initial configuration.  Examples of such models include
`groves' on the triangle \cite{CSpeyer-04,PSpeyer-05} (see also
\cite{BdTK-15} for promising results on the `massive deformation'),
and double-dimer configurations on `cubic-corner graphs' \cite{KP-16}.
Some models are in both families, for example domino tilings also
arise from the octahedron relation of cluster
algebras~\cite{Spe07,Y-09,DiFSG14}

Another important instance for phase separation and limit shape
phenomena is the six-vertex model
\cite{E-99,Zj-02,RP-10,RS-15,CGP-15}, with various choice of fixed
boundary conditions, among which the \emph{domain-wall boundary
  conditions} \cite{K-82} play a preeminent role.  The model can be
viewed as a nontrivial (`interacting', yet exactly solvable
\cite{L-67a,B-82}) generalization of the domino tilings of some domain
in the square lattice \cite{EKLP-92}.  In this context, limit shape
phenomena still need further understanding, although some progress has
been made \cite{BCG-16,BP-16} for the `stochastic' version of the model
\cite{GS-92}.

For interacting models out of free-fermionic or stochastic special
points, very few exact results are available on these phenomena.  The
evaluation of the free energy of the six-vertex model with domain-wall
boundary conditions \cite{KZj-00,Zj-00,BL-13} provided the first
quantitative indication of phase separation in the model. The sole
other result in this context concerns (a strongly supported conjecture
for) the analytic expression for the Arctic curve of the six-vertex
model on a square region of the square lattice, with domain wall
boundary conditions \cite{CP-07b,CP-07a,CP-08,CP-09,CPZj-10}. The
derivation is based on the study of an observable, the \emph{emptiness
  formation probability} (EFP), so, for short, we can call this the
\emph{EFP Method}. This result, reviewed in Section~\ref{sec.model},
shows a much richer phenomenology w.r.t.\ dimers, and more generally,
free-fermionic models: most notably the curve is \emph{algebraic} if
and only if the parameters of the model are tuned to a so-called
\emph{root of unity} case, and it is non-analytic at the points of
contact with the boundary of the domain \cite{CPN-11}.  This is at
variance with free-fermionic models, where the curve is algebraic,
and, even when non-connected, different connected components arise
naturally as different branches of the same curve \cite{KO-05}.
Because of this rich phenomenology, and with the aim of understanding
phase-separation phenomena in the presence of an interaction,
extension of these results to a larger class of domains is of great
interest.

The present paper provides an alternative approach to the EFP Method,
that we call \emph{Tangent Method}. It is based on a detailed analysis
of the line-type fundamental excitations of the frozen regions. In a
sense, it gives a `geometric interpretation' to the analytic results
arising from the EFP Method, which surprisingly had shown that the
Arctic curve is the caustic of a family of lines determined by a
single one-point boundary observable; this is here understood as the
fact that basic excitations form random walks from a given boundary
point to the Arctic curve, which are almost-straight in the
thermodynamic limit, and reach the curve tangentially, from which the
name of the method.

In this paper we use the Tangent Method to rederive the conjectured
formula for the Arctic curve in the square domain, for generic
parameters of the six-vertex models (Section \ref{sec.tangent}).  By
the same method, we determine the analytic expression for the Arctic
curve of the six-vertex model at its ice point, in a triangoloid
domain, constructed out of the crossing of three bundles of spectral
lines, that has two independent aspect-ratio parameters
(Section~\ref{sec.triangoloid}). We also provide a relation between
the Arctic curve and the generating function of the one-point boundary
correlation function, that holds for a large class of domains, and
generic values of the  parameters of the model (Section~\ref{sec.generic}).

As an instructive `minimal working example' of our method, we also
provide a very short derivation of the Arctic Ellipse for lozenge
tilings of the $a \times b \times c$ hexagon (\emph{MacMahon boxed
  plane partitions}), which is self-contained except for the use of
the Gelfand--Tsetlin formula \cite{GT-50}, thus reproducing the
classical result in \cite{CLP-98} (Appendix~\ref{app.hex}).

\section{The Arctic curve in the square: known facts}
\label{sec.model}

\subsection{The six-vertex model}

The six-vertex model is an exactly solvable model of equilibrium
statistical mechanics \cite{L-67a,B-82}.  In its simplest realisation,
it is just defined on a portion of the square grid. In its most
general `integrable' realisation, it is defined on a planar graph,
with all vertices of degree 4 and 1, obtained from the intersection
(in generic position) of a collection of open and/or closed curves in
the plane \cite{BaxPP}.

An intermediate family of domains consists of the framework of
\cite{BaxPP}, in which we have a finite number of bundles $\alpha$ of
parallel lines, pairwise mutually crossing, where the number
$n_{\alpha}$ of lines per bundle goes to infinity in the thermodynamic
limit. In this case the lattice consists of a finite number of
rectangular portions of the square lattice, glued together along some
of their boundaries.

We will consider here the most basic example of such a geometry, the
crossing of two bundles, and, in Section \ref{sec.triangoloid}, the
second simplest realisation, consisting of three bundles. In this
introduction, for sake of simplicity, we will define the model only in
the simplest case, the rectangular $N\times M$ geometry.

We have thus $NM$ vertices of degree 4, $2(N+M)$ vertices of degree 1
(\emph{external vertices}), $(N+1)M$ horizontal and $N(M+1)$ vertical
edges, of which overall $2(N+M)$ are \emph{external edges}, i.e.\ are
incident on an external vertex.  All other edges will be called
\emph{internal}.  A vertex which is first-neighbour to an external
vertex will be called a \emph{boundary vertex} (we have $2N+2M-4$ such
vertices).  An internal edge incident to a boundary vertex will be
called a \emph{boundary edge}.  We label the internal vertices with
the coordinates $(r,s)$, $r\in\{1,\dots, N\}$, $s\in\{1,\dots,M\}$, in
the obvious way.  Edges will be labeled by the coordinates of their
midpoint.

\begin{figure}
\centering
\setlength{\unitlength}{40pt}
\begin{picture}(6,2.5)
\put(0,0){\includegraphics[scale=1]{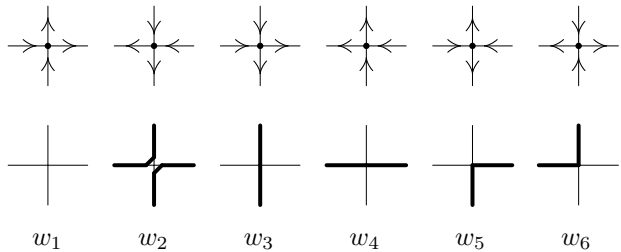}}
\put(0.5,0){\makebox[0pt][c]{$w_1$}}
\put(1.5,0){\makebox[0pt][c]{$w_2$}}
\put(2.5,0){\makebox[0pt][c]{$w_3$}}
\put(3.5,0){\makebox[0pt][c]{$w_4$}}
\put(4.5,0){\makebox[0pt][c]{$w_5$}}
\put(5.5,0){\makebox[0pt][c]{$w_6$}}
\end{picture}
\caption{The six possible types of vertex
  configurations in terms of arrows (top), or of paths (bottom), and
  their Boltzmann weights.}
\label{fig-weights}
\end{figure}

The states of the model are configurations of arrows on the edges of
the lattice, i.e.\ \emph{orientations} of the graph, satisfying the
\emph{ice rule} at all internal vertices, namely, there are two
incoming and two outgoing arrows.  This rule selects six possible
local configurations around a vertex, to which we give names as in
Fig.~\ref{fig-weights}.

An equivalent and also graphically appealing description of the
configurations of the model can be given by drawing a \emph{thick}
edge whenever an arrow is down or left, and a \emph{thin} edge
otherwise.  Due to the ice-rule, the thick edges form directed paths,
which may be oriented in such a way that all the steps are north and
east. In particular, all these paths are open and reach the boundary
of the domain.  The states of the model can thus be depicted as
configurations of paths satisfying the rules shown in
Fig.~\ref{fig-weights}.  From now on we shall mainly refer to the path
picture. Our notations are consistent with \cite[Sec.~8.3]{B-82}.

The model is further specified by assigning a Boltzmann weight $w_i$,
$i\in\{1,\dots,6\}$, to each vertex configuration, as in
Fig.~\ref{fig-weights}.  The Boltzmann weight of a given arrow configuration
$\sigma$ is the product over internal vertices
of the corresponding weight, which can be written as
\begin{equation}
\label{stateweight}
w(\sigma) =  \prod_{i=1}^6 w_i^{n_i(\sigma)},
\end{equation}
where it is understood that $n_i(\sigma)$ is the number of vertices of
type $i$ in $\sigma$. The obvious constraint $\sum_i n_i(\sigma)=NM$
shows that one such parameter is redundant.

\subsection{Fixed boundary conditions}
\label{ssec.dwbc}

Each of the external vertices may have an incoming or an outgoing
arrow. If specified in advance, we say that we have \emph{fixed
  boundary conditions}, otherwise we have \emph{free boundary
  conditions}.  Of course, a fixed boundary condition, for having any
valid configuration, requires that the overall number of incoming and
of outgoing arrows are equal. This constraint has a counterpart in
path representation: we must have as many thick edges on the south and
west sides altogether as on the north and east sides altogether.

Another useful property of fixed boundary conditions in a rectangular
geometry is that we know in advance how many horizontal and vertical
thick edges there are in the system.  Furthermore, a directed path
going (say) from the south to the north sides makes as many left-turns
as right-turns, while one going (say) from west to north makes one
more left turn. Thus, we also know in advance the difference between
the total number of left- and right-turns. This gives control on the
three linear combinations $n_5-n_6$, $2n_2+2n_4+n_5+n_6$ and
$2n_2+2n_3+n_5+n_6$. , which, together with the forementioned
$n_1+\cdots+n_6=NM$, makes only two independent parameters out of the
six weights $w_1$, \ldots, $w_6$.

For this reason, in the case of fixed boundary conditions, up to
multiplying the partition function by a trivial factor, symmetry of
the Boltzmann weights under reversal of arrows can be imposed with no
loss of generality, and it is customary to introduce the three
parameters 
\begin{align}
\fraka&:=w_1=w_2,
&
\frakb&:=w_3=w_4,
&
\frakc&:=w_5=w_6,
\end{align}
and also the convenient parameterization
\begin{align}
\label{deltat}
\Delta &= \frac{\fraka^2+\frakb^2-\frakc^2}{2\fraka\frakb},
&
t &=\frac{\frakb}{\fraka}.
\end{align}
The \emph{probabilistic space of parameters}, i.e.\ the one for which
the Boltzmann weight $w(\sigma)$ is real positive for all
configurations, is for
$\fraka, \frakb, \frakc \in \mathbb{R}^+$, and thus corresponds to 
$t \in \mathbb{R}^+$ and $\Delta < \frac{1}{2}(t + \frac{1}{t})$.

In the phase separation phenomena we will see the emergence of four
types of frozen patterns, using vertices $w_1$, \ldots, $w_4$. These
patterns can be selected on the whole domain by taking homogeneous
choices on the four sides.  Their smallest perturbation, e.g.\ having
exactly two thick edges, gives the simple problem of enumerating
configurations consisting of a single directed lattice path (discussed
in Appendix~\ref{app.directed}). Trivial as it may seem, when combined
with the more remarkable facts coming from the full-fledged six-vertex
model, this setting will prove of some use in our treatment.

Up to symmetry, we have a unique further choice of homogeneous fixed
boundary conditions, namely, in the case $N=M$, of having thick edges
on north and west sides, and thin edges on south and east sides. In
this case the geometry allows many configurations, and the paths,
which start on the west and arrive on the north being maximally
packed, travel in a rainbow fashion in the bulk, with a positive
amount of entropy. This setting is called
\emph{domain-wall boundary conditions} (DWBC) \cite{K-82}, see Figure
\ref{fig_sqsett}, left.

\subsection{Gibbs measure and correlation functions}

We now assume some fixed boundary conditions $\s_B$ have been imposed
on the boundary $\de \Lambda$ of a domain $\Lambda$.  The partition
function is the sum over the set of configurations of the model which
are compatible with the given boundary conditions, each state being
assigned its Boltzmann weight $w(\sigma)$ as in (\ref{stateweight})
\begin{equation}\label{partition}
Z_{\Lambda}(\s_B)=
\sum_{\sigma \;:\; \s|_{\de \Lambda}=\s_B} w(\sigma).
\end{equation}
Correspondingly, $w(\sigma)/Z_{\Lambda}(\s_B)$ is the Gibbs measure on the states of
the model with given boundary conditions.

For each edge $e$ of the lattice we define the characteristic
function:
\begin{equation}
\chi_{e}(\sigma):=\left\{
\begin{array}{l}
1,\quad\mathrm{if\ } e \mathrm{\ is\ thick},\\
0,\quad\mathrm{if\ } e \mathrm{\ is\ thin},
\end{array}\right.
\end{equation}
The expectation value of a product of these observables
with respect to the Gibbs measure,
\begin{equation}
\langle\chi_{e_1}, \dots, \chi_{e_n}\rangle_{\s_B}
:=
\frac{1}{Z_{\Lambda}(\s_B)}\sum_{\sigma \;:\; \s|_{\de \Lambda}=\s_B} w(\sigma)
\prod_{j=1}^{n}\chi_{e_j}(\sigma),
\end{equation}
is called an \emph{edge correlation function}. These correlation
functions clearly form a complete linear basis.  A \emph{boundary}
correlation function is a correlation function involving only boundary
edges.

\subsection{Phases of the model, and particular cases}
The study of the thermodynamic limit of the model with periodic
boundary conditions shows the emergence of three physical regimes, or
phases, according to the value of the parameter $\Delta$, namely
\emph{ferroelectric} ($\Delta>1$), \emph{anti-ferroelectric}
($\Delta<-1$), and \emph{disordered}, or \emph{critical}
($|\Delta|<1$), see \cite{B-82} for details.  In the context of phase
separation phenomena, the three phases are sometimes called
\emph{solid}, \emph{gaseous}, and \emph{liquid}, respectively.  Some
of this phenomenology survives in situations showing phase separation,
see \cite{RP-10} for details.

As anticipated, the special case $\Delta=0$ is related to free
fermions on a lattice.  In particular, at $t=1$, there is a
correspondence with non-intersecting lattice paths, dimer models and
domino tilings, the most notorious example being the so-called
\emph{domino tilings of the Aztec Diamond} \cite{EKLP-92}. In the
light of such correspondence, the model with generic value of $\Delta$
can be viewed as one of interacting dimers.  Values $t\not =1$
correspond to the presence of a non-vanishing external field (called
`bias' in \cite{JPS-98}), that favours one of the two possible
orientations of the dimers.

Another case of interest, in particular for its relations with
Algebraic Combinatorics, is the so-called \textit{ice point}, where
$\fraka=\frakb=\frakc$, and hence, $\Delta=1/2$ and $t=1$. In this
case the configurations of the model with domain wall boundary
conditions are in bijection with Alternating Sign Matrices
\cite{MRR-82,EKLP-92}, see \cite{pzjhdr} for details.

\subsection{The one-point boundary correlation function}

As a specialty of the domain-wall boundary conditions, the ice-rule
strongly constrains the pattern of vertex configurations in the
first/last row/column. For example, in the bottom-most row of vertical
edges (besides the boundary ones) there must be exactly one thick
edge, at some horizontal coordinate $1 \leq r \leq N$. We call this
value (south) \emph{refinement position}.

We will call $H_N^{(r)}$ the probability that the
refinement position is $r$, i.e., formally,
\begin{equation}
\label{HNr}
H_N^{(r)}:=\langle\chi_{e_{(r+1,3/2)}}\rangle.
\end{equation}
These quantities are naturally collected in the corresponding
generating function
\begin{equation}
\label{hNz}
h_N(z):=\sum_{r=1}^N  H_N^{(r)} z^{r-1}
=\langle\,\sum_{r=1}^N\chi_{e_{(r,3/2)}}z^{r-1}\,\rangle.
\end{equation}
This correlation function was studied, in particular, in
\cite{BPZ-02}, where it was also evaluated in the form of a
determinant.

\begin{figure}
\centering
\includegraphics[scale=1.5]{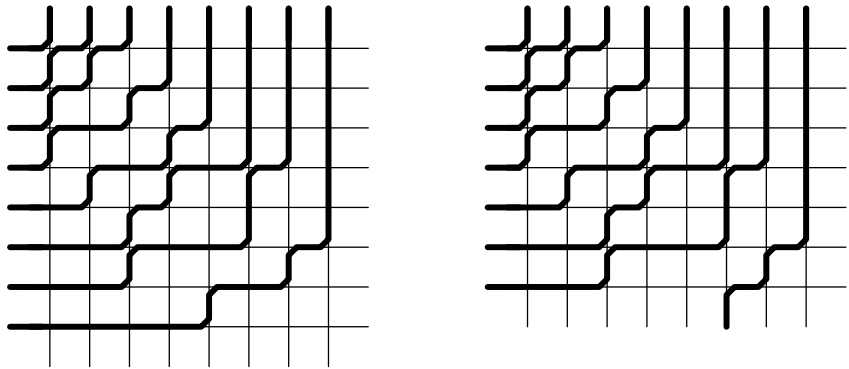}
\caption{Left, a typical configuration of the six-vertex model with
  domain-wall boundary conditions, in path representation. Here
  $N=8$. Right: a configuration in the refined ensemble with $r=6$,
  i.e.\ this configuration contributes to the probability
  $H_8^{(6)}$.}
\label{fig_sqsett}
\end{figure}

The uniqueness of the refinement position is due to DWBC, but is not
specially related to the square geometry, and holds in the more
general case of multiple bundles of spectral lines discussed above,
provided that the boundary conditions are uniform on the given side.
Below, given a domain $\Lambda$ of this form, we shall use the
notations $H_\Lambda^{(r)}$, $h_\Lambda(z)$ for the related one-point
boundary correlation function, and for the corresponding generating
function (this notation is somewhat elliptic, as it leaves understood
the precise choice of DWBC and the reference side).

\subsection{The thermodynamic limit}

The Arctic curve and limit shape phenomena are of course effects of
large volume.  At most, the liquid region of a single configuration,
taken with the Gibbs measure, is just \emph{almost surely} of the
shape given by the Arctic curve \emph{up to fluctuations}, which are
sub-linear (conjectured to be of order $N^{\frac{1}{3}}$, in analogy
with exact results in the $\Delta=0$ case \cite{J-00, J-05}, although
the precise value of this exponent, assuming that it is smaller than
1, is immaterial for the purposes of the present paper). Thus, we are
led to spend a few words on how the thermodynamic limit is performed
in presence of phase separation phenomena.  We will also introduce a
quantity, $r(z)$, which has the same information as $H_N^{(r)}$, but
will be more adapted to our purposes.

For the square domain, we just perform a thermodynamic limit $N \to
\infty$, and simultaneously rescale the lattice coordinates $(r,s)$,
in the obvious way, i.e.\ $r=\lceil Nx \rceil$, $s=\lceil Ny \rceil$,
and $(x,y)\in[0,1]^2$. We shall refer to this thermodynamic/continuum
limit as \emph{scaling limit}. In this limit, the Arctic curve is
described by an equation of the form $C(x,y)=0$. As a matter of fact,
except for the point $\Delta=0$, it is nowadays strongly believed that
this curve is analytic only piecewise, in each arc interval between
two points of contact with the boundary of the domain, thus in fact we
have, in the square, \emph{four} equations, one per corner, $C_{\rm
  SW}(x,y)=0$, and so on (for the quadruple SW, SE, NW, NE), and four
expressions for the contact points, $(0,\kappa_{\rm W})$,
$(\kappa_{\rm S},0)$ $(1,\kappa_{\rm E})$ and $(\kappa_{\rm N},0)$. Of
course, the symmetry of the problem (up to sending $t \leftrightarrow
t^{-1}$ where needed) relates the different arcs,\footnote{At $t=1$
  the curve has the obvious dihedral symmetry, and even if $t \neq 1$,
  the curve has a residual symmetry w.r.t.\ reflection along the two
  diagonals of the square. In particular
$\kappa:=\kappa_{\rm N}=\kappa_{\rm E}=1-\kappa_{\rm S}=1-\kappa_{\rm W}$.}  
and we can concentrate, say, on the south-east arc without
loss of generality (see \cite{CP-09} for more details).

If we call
\be
S_N(x)
:=
-\frac{1}{N} \ln H_N^{( \lfloor xN \rfloor )}
\ee
it is expected in general circumstances (and proven for the square
domain \cite{CP-09, CPZj-10}) that this function has a sensible limit,
i.e.\ that
$S(x) = \lim_{N \to \infty} S_N(x)$ exists, and is a convex smooth
function, with a single minimum at some $0 \leq \kappa \leq 1$ (the
contact point of the curve on this side), where it is valued zero.  In
other words, the refinement position $r$ fluctuates around its typical
value on a sub-linear range.

Similarly, the value of the function $h_N(z)$ is not quite interesting
\emph{per se}, while its derivative w.r.t.\ $z$, which allows to
extract $S(x)$ by Legendre transform, is more relevant. This suggests
to define 
\be
\label{def.rz}
r(z)
:=
  \lim_{N \to \infty}
  \frac{1}{N}z\frac{\rmd}{\rmd z}
  \ln h_N(z),
\ee
(again this limit exists and is finite for the square domain, and it
shall be in very general circumstances).  Indeed, letting
$r=\lceil\xi N\rceil$, with $0<\xi<1$, for large $N$ we may write:
\begin{equation}
\label{eq.h2H}
h_N(z) \propto \int_0^1\rmd\xi H_N^{(\lceil\xi N \rceil )} 
\rme^{\lceil\xi N\rceil\ln z}
\end{equation}
where the proportionality constant is independent of $z$, and
inessential for our purposes.  Then, from the log-convexity of $H$,
standard saddle-point arguments lead to
\begin{equation}
\label{rh1}
r(z):=\lim_{N \to \infty}
\frac{1}{N} z\frac{\rmd}{\rmd z} \ln h_N(z) = \xi_{sp}
\end{equation}
where  $\xi_{sp}$ is the solution of the  saddle-point equation
\begin{equation}
\label{rh2}
\frac{1}{N} \frac{\rmd\ }{\rmd\xi} \ln H_N^{(\lceil\xi N\rceil)} +\ln z =0.
\end{equation}
This relation will turn out to be useful below.

In the situation in which we have several crossing bundles, each
consisting of $n_{\alpha}$ lines, as we said this identifies a finite
collection of rectangles. We may define $2N=\sum_{\alpha} n_{\alpha}$,
use a local system of coordinates in each rectangle (or in each
collection of rectangles that can be glued together without conical
singularities), and rescale coordinates $(r,s)$ again in the obvious
way, i.e., under the limit $N \to \infty$ with 
$\ell_\alpha := n_{\alpha}/N$, we rescale the lattice coordinates
$(r,s)$ as $r=\lceil Nx \rceil$, $s=\lceil Ny \rceil$, and $(x,y)\in
[0,\ell_\alpha] \times [0,\ell_\beta]$ for the rectangle consisting of
the crossing of bundles $\alpha$ and $\beta$.

\subsection{The Arctic Curve Conjecture}

In \cite{CP-09}, using what we called above `the EFP method', it has
been shown that the Arctic curve of the six-vertex model on the square domain
with DWBC is completely determined by the boundary correlation
function $H_N^{(r)}$, through the following relation:

\begin{conj}[Arctic Curve Conjecture \cite{CP-09}]
\label{conj.acc}
The south-east arc of the Arctic curve of the six-vertex model with
domain wall boundary conditions can be expressed in parametric form
$x=x(z)$, $y=y(z)$, with $z \in [1,+\infty)$, as the solution of the
linear system of equations
\begin{align}
\label{FF'}
F(x,y;z) 
&= 0,
&
\frac{\rmd}{\rmd z}
F(x,y;z) 
&= 0,
\end{align}
with
\begin{equation}\label{rspe}
  F(x,y;z) = x 
- \frac{z(t^2-2\Delta t+1)}{(z-1)(t^2 z -2\Delta t +1)} y    
- r(z).
\end{equation}
\end{conj} 
\noindent
Note the change of coordinates $x\to 1-x$, with respect to
\cite{CP-09}.  The quantity $r(z)$ is defined as in (\ref{def.rz}), in
terms of $h_N(z)=h_N(z;\Delta,t)$, the generating function
\eqref{hNz}. It has a complicated (but known \cite{CP-09,CPZj-10})
expression for generic $\Delta$ and $t$, which however simplifies
considerably at the free-fermion point $\Delta=0$ (domino tilings) and
at the `combinatorial point' $(\Delta,t)=(\frac{1}{2},1)$
(alternating-sign matrices):
\begin{align}
\Delta=0\;&:
&
F(x,y;z) &= x 
- \frac{z}{z-1} \frac{t^2+1}{t^2 z+1} y
- \frac{t^2 z}{t^2 z+1}
\\
(\Delta,t)=(\smfrac{1}{2},1)\;&:
&
F(x,y;z) &= x 
- \frac{1}{z-1} y
- \frac{\sqrt{z^2-z+1}-1}{z-1}
\end{align}
We recall that by construction the solution of \eqref{FF'} provides
only one of the four portions of the Arctic curve, between two
consecutive \emph{contact points}, i.e., points where the Arctic curve
is tangent to the boundary of the square. Here we have focused on the
lower-right arc, limited by the two contact points $(1-\kappa,0)$ and
$(1,\kappa)$, corresponding to $z=1$ and $z\to \infty$,
respectively. In particular we have
\begin{equation}
1-\kappa=r(1)
=\lim_{N\to\infty}\left.\frac{1}{N} z\frac{\rmd}{\rmd z}
  \ln h_N(z)\right\vert_{z=1}.
\end{equation}
The main steps in the derivation of the result above can be summarized
as follows.

First, a specific correlation function, the \emph{emptiness formation
  probability} (EFP), devised to detect spatial transition from order
to disorder, is introduced.  This quantity, evaluated at the
coordinate $(r,s)$ on the $N\times N$ lattice, is the probability that
all the lattice sites $(r',s')$ with $r' \geq r$ and $s' \leq s$ are
occupied by a $w_1$ vertex.  Remarkably, this quantity admits an exact
formula in terms of some multiple integral
representation~\cite{CP-07b}.

Next, one has to study the asymptotic behaviour of this integral
representation in the scaling limit, in the framework of the
saddle-point approximation. In doing so, heuristic considerations
suggest to formulate a strongly supported, but still unproven
assumption that the spatial transition from order to disorder, and
hence the Arctic curve, are characterized by the condensation of
almost all roots of the saddle-point equation at the same known value.
This assumption leads directly to the Arctic Curve Conjecture
\cite{CP-07a,CP-09}.

Besides the relation between the curve and the boundary correlation
function $r(z)$, the actual determination of the expression of the
Arctic curve requires the explicit knowledge of the function $r(z)$,
and the evaluation of its behaviour in the scaling limit
\cite{CP-09,CPZj-10}.

It is worth emphasizing that, according to the conjecture above, the
form of the Arctic curve \emph{inside} the domain is completely
determined in terms of a \emph{boundary} quantity, namely the one-point
boundary correlation function.  At this stage of the reasoning,
however, there appears no clear motivation for such a relation. Our
alternative approach also addresses this point.

\section{An alternative derivation of the Arctic Curve
  Conjecture}
\label{sec.tangent}

\subsection{Preliminaries} 

To start with, let us investigate more closely in which functional
form the one-point boundary correlation function determines the Arctic
curve. It is useful to recall first some elementary geometry (see
e.g.\ \cite{librocurve}).

Let $\{ \cc_z \}_{z \in I}$ be a family of curves, in the
$(x,y)$-plane, determined by a continuous parameter $z$ valued in a
real interval $I$.  The \emph{envelope} $\eee$ of the family is the
(minimal) curve that is tangent to every curve of the family.

If the equation of the family $\{\cc_z\}$ is given in Cartesian
coordinates by $U(x,y;z)=0$, the non-singular points $(x,y)$ of the
envelope $\eee$ are the solutions of the system of equations
\begin{align}
U(x,y;z)&=0;
&
\denne{z}{{}}U(x,y;z)
&=0.
\end{align}
By analogy with caustics in geometric optics, we call 
\emph{geometric caustic} the envelope of a family of straight lines.
In this case $U$ is of degree 1 in $x$ and $y$.  This allows us to
recognise the statement of the Arctic Curve Conjecture in a compact
form: the portion of the Arctic curve is the geometric caustic of the
family of lines in the $(x,y)$-plane,
\begin{equation}
\label{eqconjbis}
U(x,y;z) = x - \frac{z(t^2-2\Delta t+1)}{(z-1)(t^2 z -2\Delta t +1)}y-r(z)
\end{equation}
for $z$ valued in the interval $[1,+\infty)$.  Note that the slope of
the lines, which is
\[
\frac{z-1}{z}
\left( 1 + \frac{t^2}{t^2-2\Delta t+1} (z-1) \right)
\]
is indeed a monotone function from $[1,+\infty)$ to 
$[0,+\infty)$, provided that 
$\Delta < \frac{1}{2}(t + \frac{1}{t})$, as is in fact implied by the
requirement of being in the probabilistic regime (i.e., 
$\fraka, \frakb, \frakc \in \mathbb{R}^+$).

This alternative formulation of the Arctic Curve Conjecture provides a
elementary geometric construction, and thus suggests the existence of
a simple `geometric principle' underlying the relation between the
Arctic curve and the boundary correlation function.

As we will explicitate in the remaining of the section, this principle
is the fact that an isolated path in a system of interacting
non-intersecting lattice paths is not sensible to the parameter
$\Delta$, thus its trajectory is locally a lattice directed random
walk, with some drift parameter fixed by the knowledge of its
endpoints.  And, in the scaling limit and at leading order, directed
paths become straight lines.  This is not unusual, as it corresponds
to the mechanism by which deterministic trajectories emerge in the
semi-classical limit of quantum theories in path-integral formulation.

\subsection{The Tangency Assumption}
\label{ssec.tangass}

\begin{figure}
\centering
\includegraphics[scale=.7]{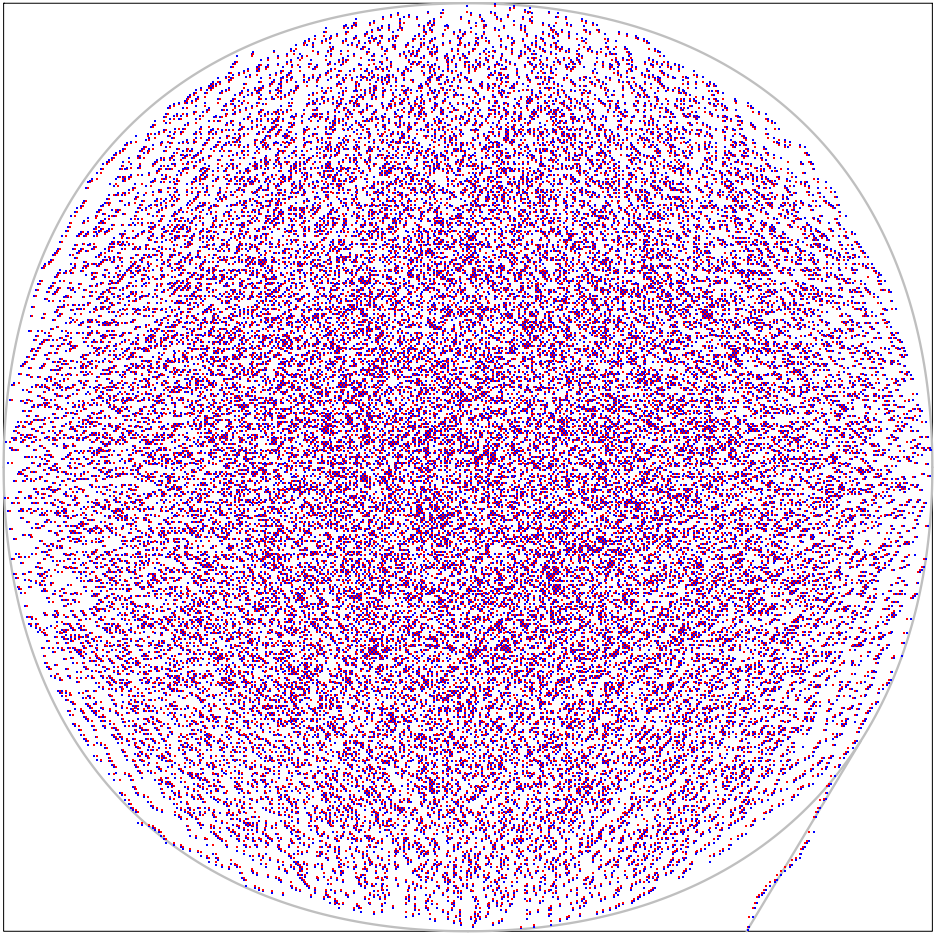}
\caption{A typical configuration of the six-vertex model, on a
  rectangular domain of size $500\times 499$, and refinement position
  $r=400$ on the south side. This configuration is exactly
  sampled,\protect\footnotemark\ at the ice point,
  $\fraka=\frakb=\frakc=1$. Blue and red dots correspond to $w_5$ and
  $w_6$ vertex configurations, respectively. In overlay in gray, the
  analytical prediction from the Tangent Assumption.}
\label{fig-refined}
\end{figure}

The quantities $H_N^{(r)}$ have been introduced as probabilities in
the model on the $N \times N$ domain, with partition function
$Z_N$. On the other side, in light of their `boundary' character, the
related quantities
$H_N^{(r)} Z_N/(\fraka^{N-r} \frakc \frakb^{r-1})$ can equally well be
seen as the partition function of a model on the
\hbox{$(N-1) \times N$} rectangle, with DWBC up to one exception: on
the south side there is one thick edge, at position $r$ (see Figure
\ref{fig_sqsett}, and Figure \ref{fig-refined} for a large example).

Yet again, the thick paths form a sort of rainbow, from their fixed
incoming positions on the south- and west-sides, to their outgoing
positions on the north-side. However this time, at difference with the
$N \times N$ case, in general they do not all start and arrive densely
packed. Two cases, depending from $r/N \lessgtr \kappa_{\rm S}$,
occur, and we shall concentrate on the case $r/N > \kappa_{\rm S}$.
Now the $N$th path, i.e.\ the only path starting from the south side,
almost surely enters the south-east frozen region, in which thick
edges are absent, and thus locally makes a directed random walk, with
some drift parameter that remains constant for a while. This behaviour
stops at the point in which the constraint of reaching the north-east
corner enters in conflict with the edge-disjointness of the thick
paths, and the presence of the liquid region inside the Arctic curve.
At this point something else must happen.  Heuristically, we expect
the path to bent, and roughly follow the profile of the Arctic curve,
up to the west contact point, and then go straight, in a frozen way,
up to its final destination endpoint.

From this heuristic scenario we are led to formulate an `assumption',
whose aim is to divide the Tangent Method in two parts. On one side,
the precise framework of the assumption provides a set of conditions
to be verified, potentially on a case-analysis to be adapted from one
system to another.  On the other side, it provides a solid basis to
establish once and for all the rigorous (but calculatory) part of the
method, which, \emph{given the assumption}, draws conclusions on the
relationship between the Arctic curve and the function $r(z)$.

\begin{assum}[Tangency Assumption]\label{assum.tangent}
  Consider the six-vertex model on the $(N-1) \times N$ domain, with
  DWBC except for the $r$th south boundary vertical edge being thick.
  In a suitable scaling limit, the resulting Arctic curve consists in
  the usual Arctic curve of the six-vertex model with domain wall
  boundary conditions, plus a straight segment, tangent to the bottom
  right portion of the Artic curve, and crossing the south boundary at
  $(r/N,0)$.
\end{assum}
\noindent
Numerical simulations strongly support the validity of this assumption
in a variety of situations, see e.g., for the ice point
$\fraka=\frakb=\frakc$, Fig.~\ref{fig-refined} and the right part of
Fig.~\ref{fig.Lshape}.

In the remaining of this section we shall summarise, somewhat in a
sketchy way, why this assumption is sounding, and which steps one
should perform in order to prove it rigorously.

\footnotetext{The numerics presented in this and other pictures has
  been generated using a {\sf C}-code based on Propp--Wilson `coupling
  from the past' algorithm \cite{PW-96}. The code, originally written
  by Matthew Blum and Jason Wolever, for the exact sampling of
  Alternating Sign Matrices, has been kindly shared by Ben Wieland; we
  have modified it to generate uniformly six-vertex model
  configurations, at ice point, on domains of various shapes.}

\begin{itemize}
\item Let us call $p=(x,y)$ the coordinate at which the $N$th path,
  starting at position $(r,0)$, first reaches a location at a distance
  $\mathcal{O}(N^{\frac{1}{2}})$ from the \hbox{$(N-1)$th} thick path.
  Then, the $N$th path, in its portion from $(r,0)$ to $(x,y)$, is
  almost surely a random (corner-weighted) directed lattice path, in
  the pertinent ensemble (as illustrated in Appendix
  \ref{app.directed}).  As such, in a large $N$ limit, it becomes a
  straight segment. This claim is completely under control.
\item Let us consider the configurations of the other $N-1$ paths,
  which determine a (rescaled) liquid region $R \subset [0,1]^2$.
  First of all, this region is expected to be almost-surely convex
  after a coarse-graining of short-scale fluctuations (this shall be
  not hard to prove). Then, \emph{conditioning} on the shape $R$, the
  position $(x,y)$ is such that the segment from $(r,0)$ to $(x,y)$ is
  \emph{tangent} to $R$, with $p$ being the tangency point. This is
  the crucial observation. If true, it must originate from the fact
  that deviations of order $N$ from the tangent trajectory, on both
  sides, would decrease the free energy. This behaviour is well under
  control provided that the interacting non-intersecting lattice paths
  ensemble has a \emph{repulsive} interaction in the frozen region of
  interest, which happens for 
  $\fraka/\frakc \leq 1$, i.e.\ $\Delta \leq t/2$.
  It is conceivable, though, that the Tangent Method, suitably
  adapted, may be applied in the full probabilistic region of
  parameters, $\fraka, \frakb, \frakc \in \mathbb{R}^+$.
\item The region $R$ concentrates, i.e., it has almost surely a given
  deterministic shape at leading order. This shall follow from the
  unicity of the associated variational problem, and thus from mild
  conditions on the form of the surface tension of the six-vertex
  model~\cite{RS-15}.
\item The deterministic limit of the region $R$ is \emph{the same} of
  the limit of the liquid region in the $N \times N$ domain. This is
  again expected, but (in our perspective) hard to formalise. Indeed,
  we have essentially peeled away part of one thick path from the
  liquid region. This makes, by itself, less volume within the region
  (but only for a sub-linear thickness), but more volume available to
  the other $N-1$ paths for drifting towards south-east, due to the
  removal of non-crossing constraints (though, less volume available
  than what would be at disposal if we peeled off \emph{the full
    path}, i.e.\ in the $(N-1) \times (N-1)$ square geometry, and we
  know that the limit shape has a thermodynamic limit). So the
  variation in the shape of $R$ is the difference of two effects, both
  sublinear, and as thus shall be sublinear.
\end{itemize}

\subsection{The domain
\phantom{aaali}}
\setlength{\unitlength}{1mm}
\begin{picture}(0.1,1)
\put(-11.5,0){$\bm{\Lambda}_{\bm{N},\bm{L}}$}
\end{picture}
\label{ssec.LampmPre}
Let us now consider yet another geometry, namely, the $N\times (N+L)$
rectangular domain, for some non-negative $L$. Let us fix the axis
origin so that the four corner vertices are located at $(1,N-1)$,
$(1,-L)$, $(N,N-1)$, $(N,-L)$. We consider the following fixed
boundary conditions: the north side has all thick edges, east and
south sides have thin edges, the west side has thick its top-most
$N-1$ edges, as well as its bottom-most, all other edges being thin.
We denote this domain, with this choice of fixed boundary conditions,
as $\Lambda_{N,L}$.

In this case, the Tangent assumption is rephrased as follows.
\begin{assum}\label{assum.lambdaLN}
  Consider the six-vertex model on the domain $\Lambda_{N,L}$. In the
  scaling limit, the Arctic curve consists in the usual Arctic curve
  of the six-vertex model with domain wall boundary condition, plus a
  straight segment, tangent to the bottom-right portion of the Artic
  curve, and reaching the south-west corner.
\end{assum}
\noindent
Indeed, the only further step from the Assumption \ref{assum.tangent}
to \ref{assum.lambdaLN} is that the straight line does not make an
angle when crossing the $N$th row, which is rather obvious from
entropic reasonings, given that the local weights are the same in the
regions above and below, and the thick path is locally far away from
other thick edges in that region.

Under Assumption \ref{assum.lambdaLN}, as we vary $L\in\mathbb{N}_0$,
in the scaling limit we obtain a family of lines in the parameter
$u=L/N$, with $u\in[0,\infty)$, that are all tangent to the
bottom-right portion of the Arctic curve. For any fixed value of $u$,
the corresponding line crosses the vertical axis at $(0,-u)$, and the
horizontal axis at some random point $(\xi,0)$, where the behaviour of
the random variable $\xi=r/N$ is discussed in a moment.  We anticipate
that this variable concentrates, so that the equation of this family
of lines in the $(x,y)$-plane is
\begin{equation}
\label{straightline}
x-\frac{\xi(u)}{u}y-\xi(u)=0,
\end{equation}
As we show below, for a certain $u=u(z)$, this is the family of
lines appearing in the geometric formulation of the Arctic Curve
Conjecture, equation~\eqref{eqconjbis}.

\subsection{Partition function of the six-vertex model on
\phantom{aaali}}
\setlength{\unitlength}{1mm}
\begin{picture}(0.1,1)
\put(-11.5,0){$\bm{\Lambda}_{\bm{N},\bm{L}}$}
\end{picture}
\label{ssec.Lampm}
As a matter of fact, once one has control on $Z_N$ and $H_N^{(r)}$,
evaluating the partition function of the model on the domain
$\Lambda_{N,L}$ is a rather easy task. For this purpose we divide the
domain $\Lambda_{N,L}$ into two portions, an upper domain
$\Lambda_{N,L}^{(+)}$, containing the top-most $N-1$ rows of vertices,
and a lower domain $\Lambda_{N,L}^{(-)}$, containing the remaining
$L+1$ rows.

For every configuration there exists one path crossing the boundary
between the two sub-domains, this occurring at some (random)
horizontal coordinate $k$. Then, the sub-domain $\Lambda_{N,L}^{(+)}$,
conditioned to this value $k$, is exactly of the form described at the
beginning of Section \ref{ssec.tangass}, and thus has partition
function
\begin{align}
\label{eq.Zpiu}
Z^{(+)}_{N,k}&:=\frac{1}{\fraka^{N-k} \frakb^{k-1} \frakc} Z_N  H_N^{(k)}
=\frac{1}{\fraka^N} \frac{1}{t^{k-1}(t^2-2\Delta t+1)^{1/2}}  Z_N  H_N^{(k)}.
\end{align}
For the sub-domain $\Lambda_{N,L}^{(-)}$ that's even simpler. The
boundary conditions have all thin edges, except for one thick edge,
$k$th from the left, on the north side, and one thick edge, the
bottom-most, on the west side. Thus we are in the situation of a
single oriented lattice path, for which the (easy) formulas are
reminded in Appendix~\ref{app.directed} in terms of the weight factors
for going straight or making a left/right turn.  The Boltzmann weights
of the six-vertex model, see Fig.~\ref{fig-weights}, induce a factor
$\frakb/\fraka$ for each straight, and $\frakc/\fraka$ for each turn.
The evaluation of the weighted enumeration of directed lattice paths
in the $y \times x$ box, a classical result in combinatorics, is
reported in equation (\ref{DPabc}).  When expressed in terms of the
quantities $\Delta$, $t$, see \eqref{deltat}, this formula reads
\begin{equation}
\label{DPdeltat}
P_{\Delta,t}(x,y)=t^{x+y+1}\sum_{l\geq 0}\binom{x}{l}\binom{y}{l} 
\left(\frac{t^2-2\Delta t+1}{t^2}\right)^{l+1/2}.
\end{equation}
Thus, for the partition function of the six-vertex model on the
lower domain, we may write:
\begin{equation}
\label{eq.Zmeno}
Z^{(-)}_{N,L,k}=\fraka^{N(L+1)}P_{\Delta,t}(k-1,L)
\end{equation}
The full partition function of the domain is then easily determined
from the combination of (\ref{eq.Zpiu}) and (\ref{eq.Zmeno})

In conclusion the partition function of the six-vertex model on the
domain $\Lambda_{N,L}$ can be expressed as
\begin{align}
Z_{\Lambda_{N,L}}&= 
\sum_{k=1}^N Z^{(+)}_{N,k} Z^{(-)}_{N,L,k}
=
\frac{\fraka^{NL} Z_N}{\sqrt{t^2-2\Delta t +1}} 
\sum_{k=1}^N 
t^{1-k}
H_N^{(k)}
P_{\Delta,t}(k-1,L)\\
&= \fraka^{NL} Z_N \sum_{k=1}^N\sum_{l\geq 0} 
\binom{k-1}{l}\binom{L}{l}  t^{L-2l} (t^2-2\Delta t +1)^{l} 
H_N^{(k)}.\label{ZNL}
\end{align}
This is an exact result, holding forn any values of $N$, $L$.  

Note that the prefactor $\fraka^{NL} Z_N$ is simply the partition
function on the same graph as $\Lambda_{N,L}$, in the case where the
$N$ thick paths start from the $N$ top-most edges of the west side
(instead that the $(N-1)$ top-most and the bottom-most), and as thus
is a useful reference normalisation.  This fact, obvious from
inspection of the possible configurations of the model in this case,
is confirmed by the above expression.  Indeed, the sum over $l$
reduces to the term $l=0$, and we are left with a sum over $k$ of
$H_N^{(k)}$, that of course evaluates to 1.

\subsection{Asymptotic behaviour of the partition function
\phantom{aaaali}}
\label{sec.sub34}
\setlength{\unitlength}{1mm}
\begin{picture}(0.1,1)
\put(-12.5,0){$\bm{Z_{\Lambda_{N,L}}}$}
\end{picture}
We now consider the expression \eqref{ZNL} in the large $N$ limit. Let
$L=\lfloor u N\rfloor$, $k=\lceil\xi N\rceil$, and $l=\lfloor\eta
N\rfloor$.  Here $\xi \in (0,1)$ and $u \in (0,+\infty)$ are rescaled
lengths, while $\eta/\xi \in (0,1)$ is a density (the fraction of
columns in $\Lambda_{N,L}^{(-)}$ with a turn). We set
\begin{equation}
\label{freeenergy}
F(u):=\lim_{N\to\infty}\frac{1}{N} 
\ln \left( \frac{Z_{\Lambda_{N,L}}}{\fraka^{NL}Z_N} \right),
\end{equation}
that is (minus) the variation in the free energy density per
horizontal step of the $N$th path, when it starts on the west side on
vertex at coordinate $(1,-L)$ rather than at $(1,0)$, as it would in
the case of ordinary domain wall boundary condition.

Note that, although $\ln Z_{\Lambda_{N,L}}=O(N^2)$ for large $N$, this
leading behaviour is completely cancelled by the term $\ln a^{NL}Z_N$.
It is easy to verify, from inspection of (\ref{ZNL}), that the limit
defined in \eqref{freeenergy} indeed exists.

The sums appearing in the expression for $Z_{\Lambda_{N,L}}$ can be
interpreted as Riemann sums, that in the scaling limit turn into a
two-dimensional real integral.  Furthermore, from the explicit
expression and the log-concavity of $H_N^{(k)}$ it is easily evinced
that the main contribution comes from a unique two-dimensional
saddle-point with positive-definite Hessian.  We are thus led to
define the `action':
\begin{align}
S(\xi,\eta;u):=&\lim_{N\to\infty}\frac{1}{N}\ln
\left[\binom{k-1}{l}\binom{L}{l}  t^{L-2l} (t^2-2\Delta t +1)^{l} 
H_N^{(k)}\right]\\
=&\ell(\xi)-\ell(\eta)-\ell(\xi-\eta)+\ell(u)-\ell(\eta)-\ell(u-\eta)
-2\eta\ln t
\\ & \qquad\qquad \qquad\qquad +\eta\ln(t^2-2\Delta t +1)
+\lim_{N\to\infty}\frac{1}{N}\ln\left[H_N^{(\xi N)}\right].
\end{align}
where we have introduced the notation $\ell(x):=x\ln x$ (adapted to
Stirling approximation). The saddle-point method gives
\begin{equation}
F(u)=S(\xi_{sp},\eta_{sp};u),
\end{equation}
where $\xi_{sp}$, $\eta_{sp}$ are the solutions of
\begin{align}
\label{SPE1}
0=\frac{\rmd\ }{\rmd\xi}S(\xi,\eta;u)&=\ln\xi-\ln(\xi-\eta)+
\lim_{N\to\infty}\frac{1}{N}\frac{\rmd\ }{\rmd\xi}
\ln\left[H_N^{(\xi N)}\right]
\\
0=\frac{\rmd\ }{\rmd\eta}S(\xi,\eta;u)&=\ln(\xi-\eta)+\ln(u-\eta)
-2\ln\eta+\ln\left(\frac{t^2-2\Delta t +1}{t^2}\right).
\label{SPE2}
\end{align}
Solving the second equation in $\eta$, one gets:
\begin{align}
\label{SPEsoleta}
\eta_{sp}
&=
\frac{1}{2\theta}\left[-(\xi+u)+\sqrt{(\xi+u)^2+4\theta\xi u}\right],
&
\theta
&:=
\frac{2\Delta t -1}{t^2-2\Delta t +1},
\end{align}
where the sign in front of the square root is fixed by requiring that
$\eta_{sp}\to 0$ as $u\to 0$, as it should, since, as we said,
$\eta_{sp}(u)$ is the average density of `turns' per horizontal
interval in the directed path from $(0,-L)$ to $(k,0)$. Note also
that, consistently, as $u$ varies over the interval $[0,\infty)$,
$\eta_{sp}$ monotonously increases over the interval $[0,1)$.

As for the first saddle-point equation, \eqref{SPE2}, by comparison
with \eqref{rh2}, its solution is just
\begin{align}
\label{SPEsolxi}
\xi_{sp}
&=r(z),
&
z
&:=
\frac{\xi_{sp}}{\xi_{sp}-\eta_{sp}}.
\end{align}
Replacing the solution \eqref{SPEsoleta} for $\eta_{sp}$ in the last
relation, and solving in $\xi_{sp}/u$ we obtain:
\begin{align}
\frac{\xi_{sp}}{u}&=\frac{z}{[(1+\theta)z-\theta](z-1)}\\
&=\frac{(t^2-2\Delta t +1)z}{(t^2 z-2\Delta t +1)(z-1)}.
\label{SPEsolxi2}
\end{align}
Plugging this last relation and expression \eqref{SPEsolxi} into
\eqref{straightline} we immediately get the statement in
Conjecture~\ref{conj.acc}, in the coordinates $x=r/N$ and $y=s/N$,
As a last consistency check, note that as $u$ varies over the interval
$[0,\infty)$, $\xi_{sp}$ monotonously increases over the interval
$[1-\kappa,1)$, and, consequently, the parameter $z$ monotonously
increases over the interval $[1,\infty)$.

\section{Extension of the method to generic domains}
\label{sec.generic}

\subsection{A local criterium}
\label{ssec.laminv}

In the derivation of the previous section, we have chosen to work in
the domain $\Lambda_{N,L}$. This is done for two reasons: for clarity
of exposition, and for matching more easily with the result of
Conjecture~\ref{conj.acc}. However, we would have obtained the same
result by adopting a variety of other families of geometries, characterised
by some volume $\Lambda_{N,L}^{(-)}$ added below the south side of the 
$(N-1) \times N$ rectangle, $\Lambda_N^{(+)}$, and with boundary
conditions so to have a single thick path starting in
$\Lambda_{N,L}^{(-)}$, within the ideas of the Tangency Assumption.

More generally, set the origin of the axes at the south-east corner of
the original square, and say that the thick path under consideration
starts from the coordinate $(-x,-y)$, with both $x$ and $y$ positive
and of order $N$. Again, in the larger $N$ limit, a portion of this
path makes a straight segment for a while. This portion is in part
contained in the square. And the crucial observation after Assumption
\ref{assum.lambdaLN}, that this straight segment does not make an
angle when crossing the boundary of the domain, still holds.

In such a geometry, up to a multiplicative factor, we would have for
the partition function a formula of the form
\be
Z_{N;x,y} \propto
\sum_r P_{\Delta,t}(x-(N-r),y) H_N^{(r)} t^{-r}
\propto
\sum_{r,l} 
\binom{x-N+r}{l}\binom{y}{l}\omega^l
H_N^{(r)}
\ee
where we use the shortcut $\w=(\frakc/\frakb)^2$, and we drop the
factors that depend on $N$, $x$ and $y$ alone.  The factor $t^{-r}$
cancels out with $t^{x-(N-r)}$ coming from $P_{\Delta,t}$, see
(\ref{DPdeltat}).  The saddle-point equations then determine $r$ and
$l$ to be concentrated on some values, the one for $r$ being most
relevant at our purposes. The equations can be obtained by comparing
the summand $(r,l)$ to $(r,l+1)$ and $(r+1,l)$, and asking for
stationarity, which gives (neglecting terms of order $1/N$)
\begin{align}
\frac{H_N^{(r+1)}}{H_N^{(r)}}
\frac{x-N+r}{x-N+r-l}
&=
1+o(1),
&
\frac{l^2}{\w (y-l) (x-N+r-l)}
&=
1+o(1).
\end{align}
Let us set $\xi=x/N$, $\eta=y/N$ and $\lam=l/N$, and recall that, from
(\ref{eq.h2H}),
the function $r(z)$ is such that
\be
\ln \frac{H_N^{(\lfloor N r(z) \rfloor+1)}}{z H_N^{(\lfloor N r(z) \rfloor)}}
=o(1).
\ee
This gives in the limit
\be
1
=
z
\frac{\xi-1+r(z)}{\xi-1+r(z)-\lam}
=
\frac{\lam^2}{\w (\eta-\lam) (\xi-1+r(z)-\lam)}
\ee
These equations are at sight homogeneous in the three independent
parameters $\eta$, $\xi-1+r(z)$ and $\lam$, which implies that the
locus of points $(-\xi,-\eta)$ (with $\eta>0$) such that the
stationary value of $r/N$ is at $r(z)$ is a straight half-line, as
expected.

Let us introduce the slope of this line, $m(z) := \eta/(\xi-1+r(z))$,
and let us change variables from $\lam$ to $d=\lam/\eta$. The
equations above become
\be
1
=
z
\frac{1}{1-md}
=
\frac{md^2}{\w (1-d)(1-md)}
\ee
from which we get, in particular, using $\w=(t^2-2 \Delta t+1)/t^2$,
\be
m(z)=
\frac{(z-1)((1-\w)z-1)}{\w z}
=
\frac{(z-1)(t^2 z -2\Delta t +1)}{z(t^2-2\Delta t+1)}
\ee
in agreement with (\ref{rspe}), once we interpret $F(x,y;z)$ as
$x-y/m(z)-r(z)$, the family (in $z$) of lines passing through
$(r(z),0)$ with slope $m(z)$.

In this formula, we got rid of the original probabilistic
interpretation, in terms of glueing of a $\Lambda^{(+)}$ original
domain and a $\Lambda^{(-)}$ accessory extra volume.  The role of the
expression (\ref{DPdeltat}) has been made completely algebraic, and
local (w.r.t.\ the neighbourhood of the one-point boundary correlation
function).  If the previous section had the goal of `geometrising' the
Arctic Curve Conjecture, the approach presented here seems paradoxally
to `de-geometrise' this very same result.  As a corollary, we can
apply our method even in geometries where, e.g.\ because of concave
angles in the domain of definition $\Lambda_N$, there seems to be no
room available for the visually-clear construction of the tangent line
in the associated extended domain $\Lambda_{N,L}$.

Alternatively, we could have imagined the extended domain
$\Lambda_{N,L}$ to live on a square lattice in a quasi-flat Riemann
surface, with a conical singularity producing the missing volume, but,
as we have seen, this is a uselessly complicated geometrical
construction for a mechanism which is algebraically clear enough.

\subsection{A more general setting}

The six-vertex model, as well as most of statistical mechanics models
for phase transitions, have been prevalently studied on some simple
domain of their underlying periodic lattice.  It is in the context of
phase separation and limit shape phenomena, where we have a strong
dependence from the boundary shape and conditions, that a study of
different domains becomes important.  Nonetheless, we have a very
modest general understanding of this feature for the six-vertex model,
especially if this is compared with the state of the art for dimer
models on bipartite lattices (see the discussion in the introduction).

\begin{figure}
\centering
\includegraphics[scale=1]{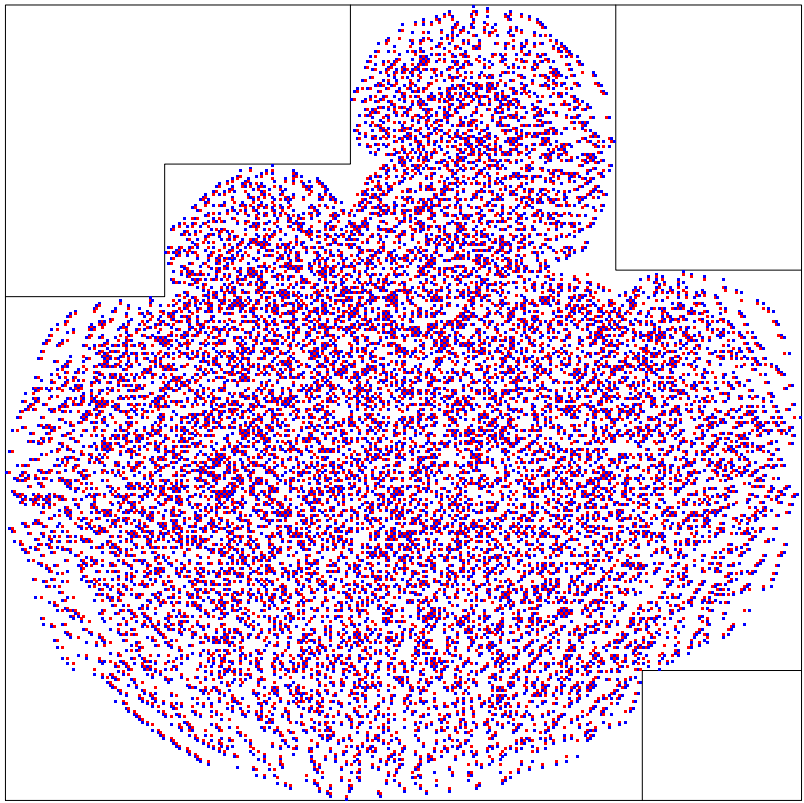}
\caption{A typical configuration of the six-vertex model at ice point,
  $\fraka=\frakb=\frakc=1$, on a generic `digitally-convex' portion of
  the square (here of size 300), with domain-wall boundary conditions.
  The side sizes, in counter-clockwise order, starting from the
  bottom-left corner, are 290, 60, 50, 70, 60, 100, 100, 70, 150, 60,
  50, 240.}
\label{fig-draft3}
\end{figure}

As we said, we believe that the most natural and appropriate context
is a version of Baxter's graphs in \cite{BaxPP}, appropriate to the
thermodynamic limit, consisting of bundles of parallel spectral lines
that mutually cross. In Section \ref{sec.triangoloid} we analyse one
such instance. A less general extension is obtained when the bundles
are divided into horizontal and vertical ones, and (as the name tells)
only horizontal and vertical bundles cross each other, determining a
portion of the square lattice which is \emph{digitally
  convex}\footnote{A digitally-convex portion of a square lattice is
  one which is enclosed by four directed paths, i.e.\ by a sequence of
  north and east steps, followed by a sequence of north and west
  steps, followed by south and west, followed by south and east,
  forming a closed non-intersecting path.}, see Figure
\ref{fig-draft3} for an example.  
We now have possibly multiple boundaries in each of the four
directions, and we say, e.g., a \emph{south side} for a horizontal
boundary of the domain, that has the domain on top of it.
The interesting case is when we have overall the same number of
horizontal and vertical lines (say, $N$), and domain-wall boundary
conditions, i.e.\ thick lines at all west and north boundaries, and
thin lines at all south and east boundaries.  We shall refer to a
setting for the six-vertex model with such a kind of domain shape and
of boundary conditions as a \emph{region of domain-wall type}.

Numerical investigations show that, analogously to what happens in the
same geometries for domino tilings, here we also have limit shapes and
Arctic curves, with the feature, new w.r.t.\ the square domain, of
having pairs of cusps in correspondence of concave angles (see Figure
\ref{fig-draft3}). Thus we have arcs of three types, connecting two
cusps, a cusp and a contact point, or two contact points. We call an
arc \emph{internal} if it is of the first type, and \emph{external} if
it is of the second or third type.

Quite evidently, these domains can be seen as marginals of the 
$N \times N$ square domain, in which the frozen regions on the four
corners have been constrained to contain the set-difference of the
square and the new domain. The simplest realisation of this,
consisting of a rectangular region cut off from the top-left corner,
just coincides with the emptiness formation probability studied in
\cite{CP-07b} and subsequent papers, so that the study of this class
of domains is promising.

The Tangent Method, in its abstraction outlined in
Section~\ref{ssec.laminv}, applies immediately to this setting, for
what concerns external arcs. Let $\Lambda$ be such a domain, let us
concentrate (say) on a given south side, and to the corner at its
right endpoint. This may be a concave corner, and is thus followed by
a west side, or a convex corner, followed by an east side. For
definiteness, we put the origin of the coordinate axes at this corner.
We aim to determine the (external) portion of the Arctic curve which
is above this south side, and on the right of its contact point (if
any), and so that the curve is `visible' from the side, i.e.\ the
tangent segment is contained within the domain. Call $H^{(r)}$, $h(z)$
and $r(z)$ the quantities associate to the one-point correlation
function pertinent to this side, in analogy with the case of a square
domain.  Let the lattice coordinates be rescaled by the same choice of
size parameter used for rescaling $r(z)$ from $h(z)$ (this may be, for
example, the total number $N$ of horizontal lines).  Then, on the same
ground of rigour of the derivation for the square domain, and based on
the suitable restating of the Assumption \ref{assum.tangent}, we have
\begin{conj}
\label{conj.gener}
  For the system outlined above, the forementioned portion of the
  Arctic curve is the geometric caustic (envelope) of the one-parameter
  family of lines in the $(x,y)$-plane, in the parameter
  $z\in[1,+\infty)$,
\begin{equation} 
F(x,y;z) = x 
- \frac{z(t^2-2\Delta t+1)}{(z-1)(t^2 -2\Delta t + z)}y
-r(z).
\end{equation}
\end{conj}
\noindent
For arcs in other orientations, we have either the very same
statement, or, if a reflection is involved, the analogous statement
with $t \leftrightarrow t^{-1}$.

\begin{figure}
\centering
\includegraphics[scale=1.3]{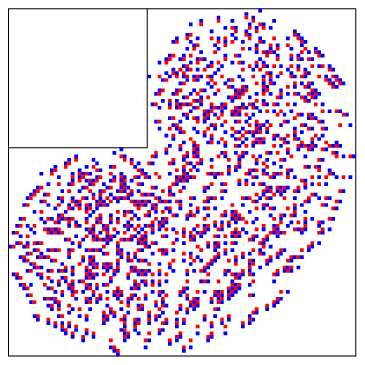}
\qquad
\includegraphics[scale=1.3]{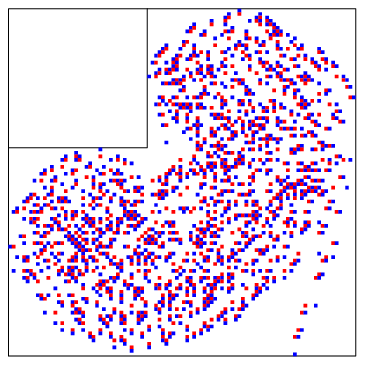}
\caption{Left: a typical configuration on the square of size 100, with
  at square of size 40 removed from the top-left corner. Right: same,
  with the extra constraint that the south refinement position is
  $r=82$. This illustrates Conjecture \ref{conj.gener}, and its use in
  the Tangent Method.}
\label{fig.Lshape}
\end{figure}

\section{The Arctic curve on the triangoloid
  domain}\label{sec.triangoloid}

\subsection{Why this model}

At this point it shall be clear that the Tangent Method applies in a
variety of circumstances. Essentially, all we need is that the model
has a conservation law in the form of line conservation, that the
behaviour of a single line is in the universality class of random
directed walks, and, apparently, that the interaction among the lines
is not of attractive type.

We have motivated already how the six-vertex model is a good prototype
for this study: it is rich enough to go beyond the free-fermionic
case, still it is probably the simplest exactly solvable model with
these characteristics, with a continuous parameter (here $\Delta$)
interpolating between universality classes. 

We have also motivated the fact that, for obtaining a sensible
thermodynamic limit, the easiest recipe is to consider a finite number
of bundles of spectra lines, that intersect each other producing
rectangular patches of the square grid, which are then arranged
together.

Then, at the light of the discussion of the previous section, one
should think that the simplest case next to treat would be the case of
Figure \ref{fig.Lshape}. Too bad that, at the moment, we are not able
to give the analytic expression of the Arctic curve for that domain.
Indeed, even assuming Conjecture \ref{conj.gener} to hold, the
quantity $r(z)$ is not known for none of the three types of sides (up
to symmetry) in this domain. What one would need in order to do so is
a fine control on some generalised version of the emptiness formation
probability (see \cite{CPS-16}).

There is a lucky situation, that we call \emph{triangoloid domain}, in
which, although the domain seems somewhat more complicated than these
other cases, at the ice point ($\fraka=\frakb=\frakc=1$) we have
access to the refined enumeration. This occurs as a corollary of the
dihedral Razumov--Stroganov correspondence \cite{CS-11}, present in
that special domain, which allows to deduce the refined enumeration
for all configurations altogether, from the known refined enumeration
on the square domain, and the one for a specially simple subclass of
configurations~\cite{CS-14}.

The determination of the Arctic curve in this domain is thus
possible,\footnote{As we say below, in a regime of aspect-ratio
  parameters there is not only an \emph{external} Arctic curve, but
  also an \emph{internal} one. We only determine the external part.}
and is the subject of this section.

\subsection{The model}\label{sec.triangoloid.1}

\begin{figure}
\centering
\setlength{\unitlength}{15pt}
\begin{picture}(16,15)
\put(0,0.2){\includegraphics[scale=1.5]{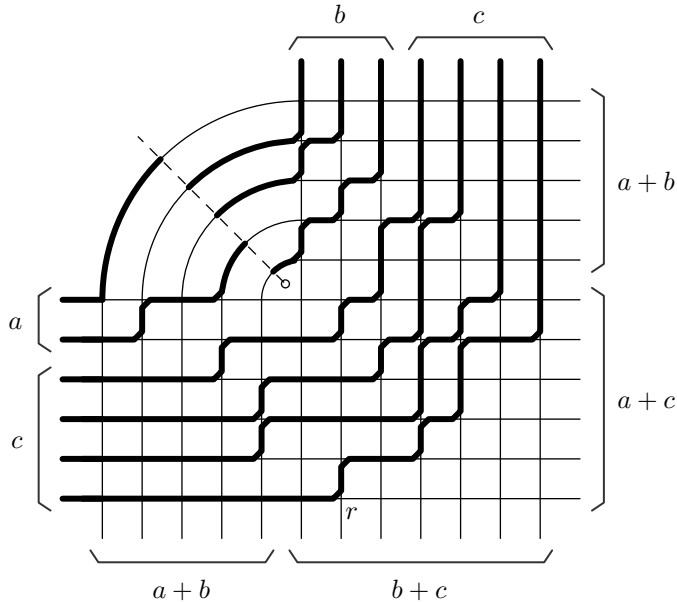}}
\put(0.5,7){\makebox[0pt][r]{$a$}}
\put(0.5,4){\makebox[0pt][r]{$c$}}
\put(15.5,5){\makebox[0pt][l]{$a+c$}}
\put(15.5,10.5){\makebox[0pt][l]{$a+b$}}
\put(4.5,0.2){\makebox[0pt][c]{$a+b$}}
\put(10.5,0.2){\makebox[0pt][c]{$b+c$}}
\put(8.5,14.7){\makebox[0pt][c]{$b$}}
\put(12,14.7){\makebox[0pt][c]{$c$}}
\put(8.8,2.2){\makebox[0pt][c]{$r$}}
\end{picture}
\caption{The $(a,b,c)$-triangoloid domain with domain wall boundary
  conditions, and a typical configuration. In this case $(a,b,c)=(2,3,4)$,
  and the south refinement position is $r=7$.}
\label{fig-triangoloid}
\end{figure}

Let $a$, $b$ and $c$ be three integers (not to be confused with the
Boltzmann weights $\fraka$, $\frakb$, and $\frakc$ of the six-vertex
model).  Take three bundles of $a+b$, $b+c$, and $c+a$ lines, crossing
each other, and use the resulting graph as a domain for the six-vertex
model, that we call \emph{triangoloid}, or \emph{three-bundle domain}
(see Fig.~\ref{fig-triangoloid}).

The internal faces of this graph are thus all squares, except for one
triangle. We have a \emph{line of defects} (denoted by a dashed line)
going from this triagular face towards the north-west corner of the
figure. Edges crossed by this line have arrows with opposite direction
on the two sides. In other words, passing to the path representation,
the two half-edges above and below the dashed line are either thick
and thin, or thin and thick, respectively.

This is a special case of six-vertex model with edge defects, whose
configurations are in fact covariant under a $\mathbb{Z}_2$ gauge in a
way analogous to frustration in two-dimensional spin glasses (this is
quickly reminded in Appendix \ref{app.gaugeZ2}), and it is useful to
keep in mind that only the endpoints of this line of defects have an
intrinsic relevance.

We take domain-wall boundary conditions, that means here that arrows
on consecutive external edges have equal orientation, unless we go
through one corner, or we go through the defect line (this,
consistently, makes a total of four changes of orientation, an even
number as it should).

For this configuration of defects, the correspondence of Figure
\ref{fig-weights} between arrows and thick-line configurations is
essentially preserved, and the thick paths are still directed, i.e.,
if oriented as outgoing from the west side and ingoing in the north
side, may only perform north and east steps.

As defects act by inverting the thickness state of the adjacent edges,
we must have an endpoint of a thick path at each and every defect.
The boundary conditions force that, of the $a+b$ defects, exactly $a$
have the endpoint of a path that started from the west side (and, in
fact, from the $a$ top-most edges of this side), and $b$ have the
endpoint of a path that terminates at the north side (and, in fact, at
the $b$ left-most edges of this side). The $c$ bottom-most edges of
the west boundary and the $c$ right-most edges of the north boundary
are connected by thick paths, that do not intersect, and pass all at
the right of the triangular face.

\begin{figure}
\centering
\includegraphics[scale=1.3]{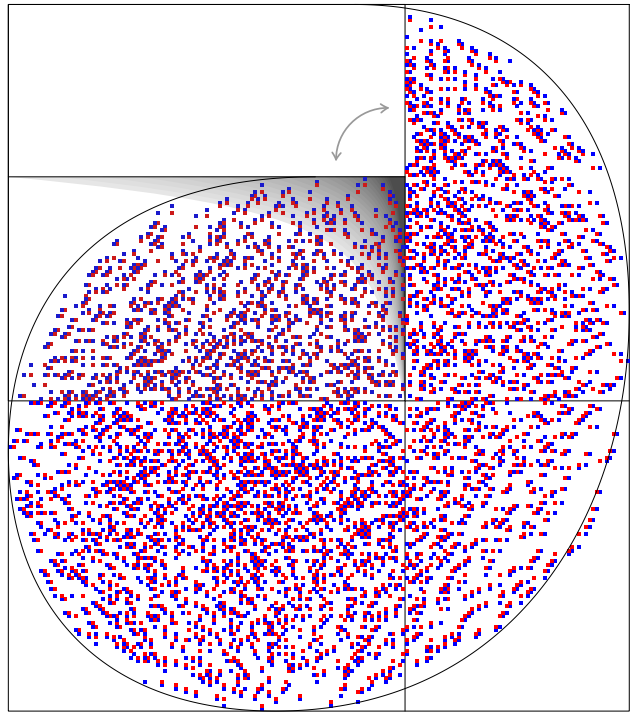}
\caption{A typical configuration of the six-vertex model at ice-point,
  $\fraka=\frakb=\frakc=1$, on a $(a,b,c)$-triangoloid. The top-right
  $(a+b) \times (b+c)$ sub-domain is reproduced a second time, rotated
  by 90 degrees, under a gray shadow, to help visualising the
  continuity of the limit shape through the line of defects.  Here
  $a=70$, $b=45$, $c=20$. In overlay, the analytic prediction of the
  Arctic curve.}
\label{fig-triangoloidshape}
\end{figure}

Numerical simulations clearly show the emergence of an Arctic curve
for triangoloids of large size. See Fig.~\ref{fig-triangoloidshape}
for an example.  Not surprisingly, in a representation showing the
collection of $\frakc$-vertices, there is no special feature occurring
at the defect line, as, in light of Appendix \ref{app.gaugeZ2}, this
line is not intrinsic to the model (it can be moved around with a
gauge transformation).

There could be, in principle, something special happening in proximity
of the triangular face, both because of the source of defects, and
because of the curvature of the square lattice at this
point. Apparently there are two regimes. When the three parameters
$a$, $b$ and $c$ are comparable, nothing special seems to happen near
to the triangle (in the case $a=b=c$ we just find back the three arcs
of ellipse for the $\Delta=1/2$ and $t=1$ square of side $2a$,
concatenated in the obvious way). When instead one parameter (say,
$c$) is small with respect to the other two, a macroscopic frozen
triangoloid region opens up around the triangular face (in the limit
$c \ll a,b$ we recover again the Arctic curve for the square domain,
this time at size $a+b$, plus a straight segment, of which we know the
coordinates from the use of the Tangent Method in Section
\ref{sec.tangent}). This second limit, which is more subtle because it
makes the Arctic curve only weakly-convex, is briefly discussed at the
end of the section.

We can adapt the construction of Section \ref{sec.tangent} to the
present case, by continuing the $a+2b+c$ vertical lines downward,
adding below the triangoloid a bundle of $L-1$ horizontal lines, and
modifying the boundary conditions as illustrated in Section
\ref{sec.tangent} (or, alternatively, we could have used the local
criterium established in Section \ref{ssec.laminv}).

As a result, the lower-right arc of the Arctic curve, delimited by the
two contact points with the right and lower boundary, may be worked
out along the lines of Conjecture \ref{conj.gener}, provided that we
can evaluate the analogue of quantity $r(z)$ for the
$(a,b,c)$-triangoloid.

\subsection{The one-point boundary correlation function at ice point}

From now on, we consider this model at ice-point,
$\fraka=\frakb=\frakc=1$, that is $\Delta=1/2$, $t=1$.

As a consequence of the ice-rule, on the south and east sides, which
are not disturbed by the defect line, there is yet again a unique
refinement position.  We will concentrate on the south one, that we
denote with $r$, and that ranges over $1 \leq r \leq a+2b+c$.

Let us denote the one-point boundary correlation
function in this case as $H_{a,b,c}^{(r)}$.  A result of \cite{CS-14} is
that, at the ice point,
\begin{multline}
\label{refined-triangoloid}
H_{a,b,c}^{(r)}=
\binom{3N-2}{N-1}^{-1}\binom{a+b+c-1}{b}^{-1}\\
\times
\sum_{s=1}^{r}
\binom{2N-s-1}{N-1}\binom{N+s-2}{N-1}
\binom{c+r-s-1}{c-1}\binom{a+b-r+s-1}{a-1},
\end{multline}
where $N:=a+b+c$, and $r\in\{1,\ldots,a+2b+c\}$.  See Appendix
\ref{app.triangoloid} for details on the genesis and derivation of
this expression.

We are interested in evaluating the asymptotic behaviour of (the
logarithmic derivative of) the corresponding generating function,
\begin{equation}
\label{habc}
h_{a,b,c}(z):=\sum_{r=1}^{a+2b+c} H_{a,b,c}^{(r)} z^{r-1}
\end{equation}
in the limit of large triangoloid sizes, $a$, $b$, $c\to\infty$, with their
ratios fixed. Let
\begin{equation}
a=\lceil N \alpha\rceil, \quad
b=\lceil N \beta\rceil, \quad
c=\lceil N \gamma\rceil, \quad
r-s=\lfloor N \xi\rfloor, \quad
s=\lceil N\eta\rceil,
\end{equation}
with $\alpha,\beta,\gamma,\xi,\eta\in\mathbb{R}$,
$\alpha,\beta,\gamma>0$, $\alpha+\beta+\gamma=1$,
$0<\xi<1+\beta$, $0<\eta<1+\beta-\eta$.

The sums appearing in \eqref{refined-triangoloid} and \eqref{habc} can
be interpreted as Riemann sums, that in the scaling limit turn into a
two-dimensional integral. Simple Stirling approximation shows the
log-concavity of the integrand, so that in the limit the integral
is dominated by the contribution of a unique non-singular
saddle point. We define the `action':
\begin{multline}
S(\xi,\eta;\alpha,\beta,\gamma,z):=\\
\lim_{N\to\infty}\frac{1}{N}\ln \left[
\frac{(2N-s-1)!(N+s-2)!(c+r-s-1)!(a+b-r+s-1)!}{(N-s)!(s-1)!(r-s)!(b-r+s)!}
z^{r-1}\right],
\end{multline}
where we have ignored factors that do not depend on $\xi$ or $\eta$.

From standard saddle-point arguments, it follows that
\be
\begin{split}
\label{rabc}
r_{\alpha,\beta,\gamma}(z)&:=
\lim_{N\to\infty}\frac{1}{N}z\frac{\rmd}{\rmd z}\ln h_{a,b,c}(z)\\
&=\xi_{sp}+\eta_{sp},
\end{split}
\ee
where $\xi_{sp}$, $\eta_{sp}$, are the solutions of the two
saddle-point equations:
\begin{align}
z&=\frac{\xi(1-\gamma-\xi)}{(\beta-\xi)(\gamma+\xi)},
&
z&=\frac{\eta(2-\eta)}{1-\eta^2},
\end{align}
namely,
\begin{subequations}
\label{trispesol}
\begin{align}
\xi_{sp}&=\frac{(\gamma-\beta)z+\alpha+\beta-    
\sqrt{[(\gamma-\beta)z+\alpha+\beta]^2-4\gamma\beta z(1-z)}} {2(1-z)}
\\
\eta_{sp}&=\frac{1-\sqrt{z^2-z+1}}{1-z}
\end{align}
\end{subequations}
The signs of the square roots are fixed by the condition that for
large real $z$ the quantity $r_{\alpha,\beta,\gamma}(z)$ should tend
to the value:
\begin{equation}
\lim_{z\to+\infty}(\xi_{sp}+\eta_{sp})=
\frac{\alpha+2\beta+\gamma}{\alpha+\beta+\gamma}=
1+\beta.
\end{equation}
This condition follows directly from definitions \eqref{habc} and
\eqref{rabc}, assuming that the limits $N\to\infty$ and $z\to\infty$
may be interchanged.

\subsection{The Arctic curve for the six-vertex model on the
  triangoloid}

We are now ready to use Conjecture \ref{conj.gener}.  Setting
$\Delta=1/2$, $t=1$ in \eqref{eqconjbis}, and inserting the quantity
$r_{\alpha,\beta,\gamma}(z)$, see \eqref{rabc}, expressed as the sum
of the solutions \eqref{trispesol}, we obtain the family of lines
\begin{multline}
y=(z-1)x+1-\sqrt{z^2-z+1}+
\frac{1}{2}\left[(\gamma-\beta)z+\alpha+\beta\right]\\
-\frac{1}{2}    
\sqrt{[(\gamma-\beta)z+\alpha+\beta]^2 + 4\gamma\beta z(z-1)},
\qquad z\in[1,\infty).
\end{multline}  
The corresponding geometric caustic has the parametric form
\begin{equation}\label{tri.arctic1}
\left\{
\begin{array}{l}
x=1+\beta-\zeta(z;\alpha,\beta,\gamma)
\\
\\
y=\zeta(\frac{z}{z-1};\beta,\alpha,\gamma)  
\end{array}
\right.\qquad z\in[1,\infty)
\end{equation}
where
\begin{equation}
\label{tri.arctic2}
\zeta(z;\alpha,\beta,\gamma)
=\frac{3-\alpha}{2}-\frac{2z-1}{2\sqrt{z^2-z+1}}
-\frac{(1-\alpha)^2 z+\alpha\gamma-\beta}{
2\sqrt{[(\gamma-\beta)z+\alpha+\beta]^2-4\gamma\beta z(1-z)}}
\end{equation}
This describes the south-east arc of the Arctic curve of the
six-vertex model at ice-point, on the triangoloid, between the two
contact points on the east and south boundaries.  As evinced from the
$z=1$ limit of the expression above, and more easily from
(\ref{refined-triangoloid}), the south contact point is at the
rescaled coordinate $(\kappa,0)$, with 
$\kappa = \frac{\alpha+\beta+\gamma}{2}+\frac{\beta
  \gamma}{\alpha+\gamma}$.

\begin{figure}[!tb]
\centering
\includegraphics[scale=1.3]{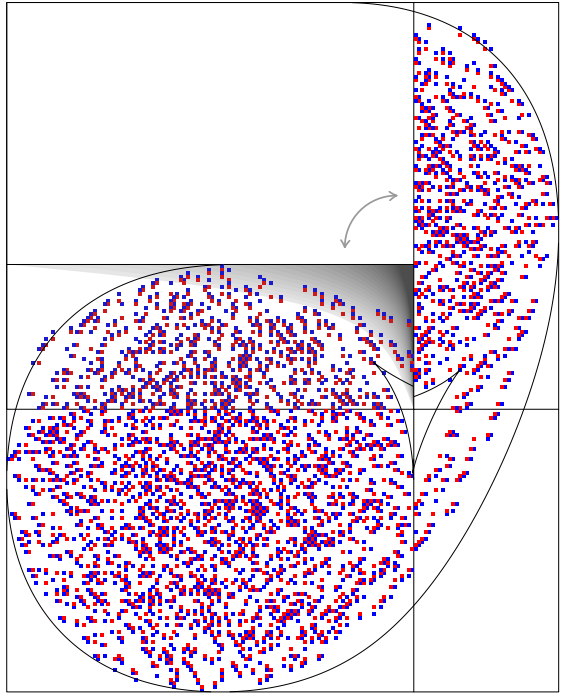}
\caption{Analogue of Figure \ref{fig-triangoloidshape} for a different
  aspect ratio. In this case $a=79$, $b=39$, $c=3$. In overlay, the
  analytic prediction of the external portion of the Arctic curve, and
  the curve obtained by the `wild guess' described in the text, which
  reproduces the internal portion of the Arctic curve in the limit
  $\gamma \to 0$.}
\label{fig-triangoloidshapeBuco}
\end{figure}

The other two arcs of the curve can be obtained straighforwardly by
cyclic permutation of the parameters $\alpha$, $\beta$, $\gamma$ in
\eqref{tri.arctic1}, \eqref{tri.arctic2}, and appropriate relabeling
of coordinate axis.  The result is plotted against a numerical
simulation in Fig.~\ref{fig-triangoloidshape}.

We have already mentioned that, in the limit $c \ll a,b$, we shall
obtain back the Arctic curve of the square, reproduced within the
south-west and north-east rectangular sub-domains w.r.t.\ Figure
\ref{fig-triangoloid} (cut towards the line of defects), plus a
straight segment, on the south-east side, tangent to both copies of
the Arctic curve (the numerical simulation of Figure
\ref{fig-triangoloidshapeBuco} is not far from this limit).  This may
seem mysterious at first, as, for positive values of the size
parameters, the resulting curve is convex.

In order to see how this limit develops a singularity, consider the
expression (\ref{tri.arctic2}) for $\gamma \to 0^+$ (and thus
$\alpha+\beta \to 1^-$). In this limit we have
\begin{equation}
\label{tri.arctic2mod}
\zeta(z;1-\beta,\beta,0)
=1+\frac{\beta}{2}
-\frac{2z-1}{2\sqrt{z^2-z+1}}
+
\frac{\beta}{2}
\frac{1-\beta z}{\sqrt{(1 - \beta z)^2}}
\end{equation}
and the quantity $\frac{1-\beta z}{\sqrt{(1 - \beta z)^2}}$ has to be
interpreted as the sign of $1-\beta z$. Thus the $x(z)$ coordinate
function has a jump for $z=1/\beta$, and, consistently, $y(z)$ has a
jump for $z/(z-1)=1/\alpha=1/(1-\beta)$, i.e.\ again for $z=1/\beta$.

In fact, in the limit $\gamma \to 0$, if in both entries of the
parametric solution (\ref{tri.arctic1}) we use a function $\zeta$ with
\emph{the other} sign of square root in the last summand, w.r.t.\ the
definition in (\ref{tri.arctic2}), we obtain a curve that approaches the
known \emph{internal part} of the Arctic curve (see Figure
\ref{fig-triangoloidshapeBuco}). However, this wild guess, besides
being not theoretically motivated, must also be wrong in some respect
when $\gamma$ is small but positive. In this case the curve has the
appropriate qualitative behaviour (including the cusps, and some of
the consistency checks), \emph{but} the two endpoints of the curve,
when folded around the conical singularity, miss each other by a
distance $\mathcal{O}(\gamma)$, namely
$\gamma \frac{1-\alpha \beta}{(\alpha + \gamma)(\beta + \gamma)}$.

\section{Conclusions}

\subsection*{Which models next?}
This paper sets the basis for a method aimed at the determination of
the Arctic curve in statistical mechanics models on planar graphs,
with (piecewise) local translational invariance of the lattice and
weights, showing phase separation phenomena, in light of a conserved
quantity in the associated transfer matrix, that can be seen as the
number of lines in a suitable line representation.
This is the case for a variety of dimer or free-fermionic models (for
which, however, in most cases more powerful general methods already
exist), for the six-vertex model, treated here in detail, and for
variants of it in which some spectral lines may contain higher spin or
$q$-bosons.

The constraint of having a line representation may appear as a strong
limitation of the method. Let us however stress how, up to bijections,
families of non-intersecting (possibly interacting) lattice paths
constitute a very general and flexible language for representing a
variety of mathematical structures, ranging from Young diagrams
\cite{RS-16} and tableaux \cite{PR-07} to $Q$-systems and cluster
algebras \cite{DFK-10}.

The application of the Tangent Method to quite different classes of
models, and the study of its interplay with other existing techniques,
is in our opinion a direction of research deserving to be explored.

\subsection*{What more for the six-vertex model?}

It shall be clear that also for the six-vertex model, the main subject
of this paper, the analysis is far from complete. A variety of domains
still asks for the determination of their Arctic curve, and in
particular it would be quite interesting to obtain the analytic
expression for an Arctic curve in a domain presenting cusps, for a
system out of free-fermionic points.  As we said, the main obstacle in
these derivations is the lack of knowledge of refined enumerations,
called here one-point boundary correlation functions.  The most
promising candidate seems to be the six-vertex model in the domain
presented in Figure \ref{fig.Lshape}.  We hope that some progress in
this direction will be available in the light of our results on a
generalisation of the Emptiness Formation Probability
observable~\cite{CPS-16}.

We also remind that the Tangent Method, in its present formulation is
not adapted to the determination of the internal portions of the
Arctic curve, i.e.\ the arcs between two cusps. Or the internal
components of Arctic curves in the cases where, as for the triangoloid
domain, there are internal frozen regions. We hope that the puzzling
features of the internal component of the curve outlined at the end of
Section \ref{sec.triangoloid} may be clarified in the future.

\subsection*{Which Tangent Method?}

Another natural question is how to make precise the assumptions listed
in Section \ref{ssec.tangass}. In principle, the short discussion
following the assumption gives a clear roadmap to this task. However,
a further aspect of the method that we have not discussed here is the
fact that it exists in several variants, which exploit in slightly
different ways the peculiar behaviour of one thick path that, in one
way or another, has been singled out from the liquid region by mean of
a marginalisation on a boundary observable.

The version described in this paper, that could be called
\emph{Geometric Tangent Method}, is the one which is visually more
clear (especially in its realisation with an auxiliary external
domain, as in Section \ref{sec.tangent}). However, we have devised
also an \emph{Entropic Tangent Method}, that establishes a criterium
based on the locality of the free energy, and performs some `surgery'
of domains for comparing the free energy of different refined
ensembles.  We have an \emph{Algorithmic Tangent Method}, adapted to
those cases in which the configurations are obtained from iterated
applications of substitutional rules. Finally, we have a promising
\emph{Doubly-refined Tangent Method}, which exploits (when available)
the two-point boundary correlation function, the two points being on
two consecutive sides of the boundary, this providing a geometric
setting in which the complicancy of the contact between the tangent
path and the liquid region is eliminated.

All these different methods come with slightly different technical
requirements, for satifying the associated variants of the Tangent
Assumption, and the comparison between the different methods is still
to be completely investigated, in a trade off between the mathematical
control on the assumptions, and the domain of applications.

\section*{Acknowledgments}

\noindent
We are indebted to Luigi Cantini and Andrei Pronko for useful
discussions.  We are grateful to Ben Wieland for sharing with us the
code for generating uniformly sampled Alternating Sign Matrices.  We
thank the Mathematical Science Research Institute (MSRI, Berkeley),
research program on `Random Spatial Processes', the Simons Center
for Geometry and Physics (SCGP, Stony Brook), research programs on
`Conformal Geometry' and on `Statistical Mechanics and
Combinatorics', the Institute for Computational and Experimental
Research in Mathematics (ICERM, Brown University), research program on
`Phase Transitions and Emergent Properties', and the Galileo Galilei
Institute for Theoretical Physics (GGI, Florence), research program on
`Statistical Mechanics, Integrability and Combinatorics') for
hospitality and support at some stage of this work.  FC is grateful to
LIPN/Equipe Calin, and AS is grateful to INFN, Sezione di Firenze, for
hospitality and support during part of this work.

\appendix

\section{Weighted enumeration of directed lattice
  paths}\label{app.directed}

\noindent
We recall here some classical results in analytic combinatorics,
concerning the enumeration of two-dimensional directed lattice paths
in the square lattice, weighted according to the number of `corners',
and recast them in a form suitable for our purposes.

A directed lattice path $\gamma:(0,0) \to (x,y)$ is a path on the
square lattice, starting in $(0,0)$ and arriving in $(x,y)$, and whose
only allowed steps are $(1,0)$, or `east', and $(0,1)$, or `north'. If
the path visits the vertices $\{v_i\}_{0 \leq i \leq x+y}$, we thus
have
$v_{i+1}-v_i \in \{(1,0),(0,1)\}$, $v_0=(0,0)$ and $v_{x+y}=(x,y)$.

For $x$ and $y$ nonnegative integers, the number of such paths
reaching the point of coordinates $(x,y)$ is clearly
\begin{equation}
\label{DP}
P(x,y)= \binom{x+y}{y}.
\end{equation}
We now assign to each path $\gamma$ a weight
$\omega^{c(\gamma)}$, where $c(\gamma)$ is the
number of `north-east corners', i.e., of vertices $v_i$, $0< i< x+y$
preceeded and followed by a north- and an east-step, respectively,
$v_i = v_{i-1} + (0,1) = v_{i+1}-(1,0)$.
We want to evaluate the weighted enumeration
\begin{equation}
P_\omega(x,y):=\sum_{\gamma} \omega^{c(\gamma)}.
\end{equation}
Let $\mathcal{N}(x,y,l)$ denote the number of directed lattice paths
reaching $(x,y)$, with exactly $l$ north-east corners. It is clear
that these paths are in bijection with pairs of subsets of 
$I \subseteq \{0,\ldots,x-1\}$ and $J \subseteq \{1,\ldots,y\}$, both
of cardinality $l$ (the $k$th corner is at $v=(i_k,j_k)$, where $i_k$
and $j_k$ are the $k$th element of sets $I$ and $J$, in order).
This leads immediately to
\begin{equation}
\mathcal{N}(x,y,l)= \binom{x}{l}\binom{y}{l},
\end{equation} 
and, hence,
\begin{equation}
\label{DPomega}
P_\omega(x,y)=\sum_{l\geq 0}  \binom{x}{l}\binom{y}{l}\omega^l.
\end{equation} 
Obviously, in the above sum all terms with $l>\mathrm{min}\{x,y\}$
vanish. Note also that the formula above consistenlty reduces to
\eqref{DP} at $\omega=1$, as a result of Chu-Vandermonde formula.

For our purposes, it is now convenient to slightly modify our
definition by adding to each lattice path $\gamma$ an east step just
before the origin, and a north step next to the final point $(x,y)$.
We denote this modified path by $\tilde{\gamma}$. Manifestly, the new
path does not have any extra north-east corner.

We now want to count paths $\tilde{\gamma}$, according to two
statistics: number $s(\tilde{\gamma})$ of `straights', i.e., of
vertices $v_i$, $0\leq i \leq x+y$, such that the preceeding and
following steps are both north or both east, and the number
$t(\tilde{\gamma})$ of `turns', i.e., of vertices $v_i$, $0\leq i \leq
x+y$, such that the preceeding and following steps are either north
and east, or east and north.  It is clear that
\begin{align}\label{app_rel}
&s(\tilde{\gamma}) + t(\tilde{\gamma}) =  x+y+1, \\
&t(\tilde{\gamma}) = 2 c(\tilde{\gamma})+1.
\end{align}
which makes manifest the homogeneity of this double-statistics, and
its connection with the previous formula (\ref{DPomega})

Let us now assign weights $\frakb/\fraka$ and $\frakc/\fraka$ to each
straight and turn, respectively. For what we said, the corresponding
weighted enumeration of paths
\begin{equation}
P_{\fraka,\frakb,\frakc}(x,y):=\sum_{\tilde{\gamma}} 
\Big(\frac{\frakb}{\fraka}\Big)^{s(\tilde{\gamma})} 
\Big(\frac{\frakc}{\fraka}\Big)^{t(\tilde{\gamma})}.
\end{equation}
is just given by
\begin{align}\label{DPabc}
P_{\fraka,\frakb,\frakc}(x,y)&=\sum_{l\geq 0} \mathcal{N}(x,y,l) 
\Big(\frac{\frakb}{\fraka}\Big)^{x+y-2l} 
\Big(\frac{\frakc}{\fraka}\Big)^{2l+1}\\
&=\Big(\frac{\frakb}{\fraka}\Big)^{x+y+1}
\sum_{l\geq 0}\binom{x}{l}\binom{y}{l} 
\Big(\frac{\frakc}{\frakb}\Big)^{2l+1}.
\end{align}
Again, in the above sum all terms with $l>\mathrm{min}\{x,y\}$ vanish.
This formula can be applied directly in the context of the six-vertex
model, with Boltzmann weights $\fraka$, $\frakb$, $\frakc$ as in
Section \ref{ssec.dwbc}.  Use of \eqref{deltat} to express the
Boltzmann weights in terms of the parameters $\Delta$ and $t$ leads to
equation~\eqref{DPdeltat}.

\section{Alternating Sign Matrices}
\label{app.asm}

\noindent
An Alternating Sign Matrix (ASM) of size $n$ is an $n\times n$ matrix
valued in $\{0,\pm 1\}$, such that: (i) non-zero entries alternate in
sign along rows and columns; (ii) the sum of entries along each row or
column is $+1$ \cite{pzjhdr}.

Let $A_n$ be the number of ASMs of size $n$. It is well known that
\cite{Ze-96,Ku-96}:
\begin{equation}\label{ASMenum}
  A_n=\prod_{j=0}^{n-1}\frac{(3j+1)!}{(n+j)!}.
\end{equation}
ASM of sixe $n$ are in bijection with the configurations of the
six-vertex model on the $n\times n$ lattice, with domain wall boundary
conditions (entries $+1$ and $-1$ in the ASM corresponding to $w_6$
and $w_5$ vertices in the six-vertex model, respectively).  Thus the partition
function of the model \cite{I-87}, when evaluated at the ice point,
$\fraka=\frakb=\frakc=1$, coincides with $A_n$.

The one-point correlation function $H_N^{(r)}$, then, is related to
the so-called \emph{refined enumerations}: Let $A_n(r)$ be the number
of ASM of size $n$ such that the sole non-zero entry in the bottom row
is in the $r$th column, then it is well-known that~\cite{Ze-96b}:
\begin{equation}
\label{ASMref}
  A_n(r)=A_n \binom{2n-r-1}{n-1} 
\binom{n+r-2}{n-1} \binom{3n-2}{n-1}^{-1}.
\end{equation}
and it is clear that $H_n^{(r)}|_{\Delta=\frac{1}{2}, t=1} = A_n(r)/A_n$.

Then, in the formalism of Section \ref{ssec.Lampm}, the partition
functions $ Z^{(+)}_{N,k} $ and $ Z^{(-)}_{N,L,k} $ just reduce to
$A_N(k)$, as in \eqref{ASMref} above, and to the binomial coefficient
$P(r,L)$, as in \eqref{DP}. In particular, one can calculate
\begin{equation}
r_{\rm ASM}(z):=\lim_{N\to\infty}\frac{1}{N}z\frac{\rmd}{\rmd z}
\ln \left(\frac{1}{A_N}\sum_{r=1}^N A_N(r) z^{r-1}\right)=
\frac{\sqrt{z^2-z+1}-1}{z-1},
\end{equation}
which gives the family of lines
\begin{equation}
F_{\rm ASM}(x,y;z) = 
x-\frac{1}{z-1}y-r_{\rm ASM}(z),    \qquad z\in[1,+\infty).
\end{equation}
The corresponding geometric caustic reproduces (the lower-right quarter
of) the limit shape of ASMs, first derived in~\cite{CP-08}.

\section{The Arctic curve for lozenge tilings of a
  hexagon}
\label{app.hex}

\noindent
As a simple application of the Tangent Method, in a framework
different from the six-vertex model, we derive here the Arctic curve
for lozenge tilings of a hexagon, or equivalently, the frozen boundary
of the limit shape of boxed plane partitions, which is an ellipse,
thus recovering a result of Cohn, Larsen and Propp \cite{CLP-98}.

Let us call a \emph{$(a,b,c)$-hexagon} the hexagonal portion of the
triangular lattice whose side lengths are $a,b,c,a,b,c$ in clockwise
order, and let us adopt the convention that the horizontal sides are
those of length $b$.

We are interested in the tilings of this hexagon with rhombi of side
length 1, i.e.\ obtained from the union of two neighbouring triangles
of the lattice, called \emph{lozenges}.  Referring to the orientation
of the longest diagonal, we have one type of \emph{vertical} lozenges,
and two orientations for \emph{oblique} lozenges.  In this Appendix we
shall adapt to the notations of \cite{CLP-98}.

Let $M_{a,b,c}$ be the number of lozenge tilings of a
$(a,b,c)$-hexagon. We recall that
\begin{equation}\label{macmahon}
M_{a,b,c}
=\prod_{j=0}^{a-1}\prod_{k=0}^{b-1}\prod_{l=0}^{c-1}\frac{j+k+l+2}{j+k+l+1}
=\prod_{j=0}^{b-1}\frac{j!(j+a+c)!}{(j+a)!(j+c)!},
\end{equation}
a classical result of MacMahon.

In order to apply the Tangent Method, we need some particular refinement of
MacMahon formula. These very same quantities have already been evaluated in
\cite{EKLP-92,CLP-98}, exploiting a nice description of the lozenge
tilings in terms of \emph{semi-strict Gelfand--Tsetlin patterns}
\cite{GT-50}, which we now recall.

\begin{figure}[!tb]
\begin{center}
\setlength{\unitlength}{17.67pt}
\begin{picture}(11,5)(0.15,0.7)
\put(0.02,0.7){\includegraphics[scale=1.7]{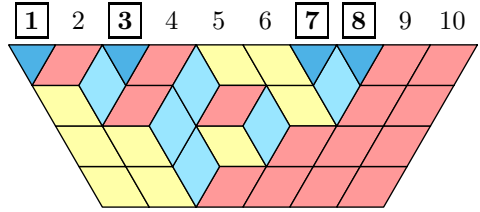}}
\put(1,5){\makebox[0pt][c]{\framebox{\bf 1}}}
\put(2,5){\makebox[0pt][c]{2}}
\put(3,5){\makebox[0pt][c]{\framebox{\bf 3}}}
\put(4,5){\makebox[0pt][c]{4}}
\put(5,5){\makebox[0pt][c]{5}}
\put(6,5){\makebox[0pt][c]{6}}
\put(7,5){\makebox[0pt][c]{\framebox{\bf 7}}}
\put(8,5){\makebox[0pt][c]{\framebox{\bf 8}}}
\put(9,5){\makebox[0pt][c]{9}}
\put(10,5){\makebox[0pt][c]{10}}
\end{picture}
\end{center}
\caption{\label{fig-gelfand}A lozenge tiling of a $(n,k)$-trapezoid as
  described in the text, with $(n,k)=(4,6)$ and ${\bm x}=(1,3,7,8)$.}
\end{figure}

Define the \emph{$(n,k)$-trapezoid} as the isosceles trapezoid region
of the triangular lattice, with sides $k,n,n+k,n$, in order. We will
represent this with the short basis on the bottom, and the long basis
on top. Tilings of this region with lozenges and unit triangles that
maximise the number of lozenges have exactly $n$ triangles. We will
consider such tilings, under the restriction that all these triangles
are adjacent to the long basis.  For a given tiling, we denote by
$\mathbf{x}=(x_1,\dots,x_n)$, with 
$1\leq x_1<x_2<\dots<x_{n-1}<x_n\leq n+k$, the horizontal coordinates
of these triangles.  See Fig.~\ref{fig-gelfand} for an example.  Each
tiling of the above defined trapezoid can be equivalently viewed as a
semi-strict Gelfand--Tsetlin pattern with top row $\mathbf{x}$, see,
e.g., \cite{CLP-98}.

Let $V_n(\mathbf{x})$ be the number of tilings of the
$(n,k)$-trapezoid with triangular tiles located at $\mathbf{x}$. A
well-known formula, due to Gelfand and Tsetlin, states
\cite{GT-50,EKLP-92,CLP-98}:
\begin{equation}\label{GT}
V_n(\mathbf{x})=\prod_{1\leq i<j\leq n} \frac{x_j-x_i}{j-i}.
\end{equation}
The number of tilings does not depend on $k$, as long as $k>x_n-n$,
since for non-minimal values of $k$ there is a frozen region on the
right side. Similarly, it is invariant under an overall translation
$x_j\to x_j+ l$  because of a frozen region on the left side.

It is easy to see that, under the choice $k=b$, $n=a+c$,
$\mathbf{x}=(1,\dots, a,a+b+1,\dots,a+b+c)$, two frozen equilateral
triangles, of side $a$ and $c$, appear in the two upper corners of the
trapezoid, and what is left to tile (with lozenges only) is exactly a
$(a,b,c)$-hexagon.  Indeed, in this case, the Gelfand--Tsetlin formula
\eqref{GT} reduces, after some manipulations, to MacMahon formula,
equation \eqref{macmahon}. We shall now consider two other
specializations of the Gelfand--Tsetlin formula, that lead to refined
enumerations of lozenge tilings of a hexagon (i.e., in the language of
this paper, to one-point boundary correlation function), and are thus
adapted to the application of the Tangent Method.

The first case corresponds to the particular choice\footnote{As
  customary, $\ldots, \hat{r}, \ldots$ stands for $\ldots, r-1,r+1,
  \ldots$.}  $\mathbf{x}=(1,\dots,\hat{r},\dots$,
$a+1,a+b+1,\dots,a+b+c)$.  With respect to the MacMahon realisation,
one triangle has been moved from $r$ to $a+1$.  We shall denote the
number of tilings of this region as $M_{a,b,c}(r)$, which is just the
shortcut
\be
M_{a,b,c}(r)=V_{a+c}(1,\dots,\hat{r},\dots,a+1,a+b+1,\dots,a+b+c).
\ee
Clearly we have the extremal values
\begin{align}
M_{a,b,c}(a+1)&=M_{a,b,c},
&
M_{a,b,c}(1)&=M_{a,b-1,c}.
\end{align}
In working out explicit expressions, it is significantly simpler to
evaluate ratios. In the present case, an elementary calculation gives:
\begin{equation}\label{hexref1}
\frac{M_{a,b,c}(r)}{M_{a,b,c}}
\equiv
\frac{M_{a,b,c}(r)}{M_{a,b,c}(a+1)}=
\binom{a}{r-1}\binom{b+c-1}{c}\binom{a+b+c-r}{c}^{-1},
\end{equation}
with $r\in\{1,\dots,a+1\}$.

The second case of interest is the enumeration of the lozenge tilings
of the $(a,b,c)$-hexagon, refined according to the location of the
unique vertical lozenge occuring in the vicinity of the top boundary.
Cutting away the top-most row of the lattice, this can be rephrased as
the enumeration of lozenge tilings of a trapezoid with bases $b$ and
$a+b+c-1$, heigth $a+c-1$, and triangular tiles located at
$\mathbf{x}= (1,\dots,a-1,a+r,a+b+1,\dots,a+b+c-1)$. Let us denote
this number as $N_{a,b,c}(r)$, where $r\in\{0,1,\dots,b\}$, which is
just the shortcut \be
N_{a,b,c}(r)=V_{a+c-1}(1,\dots,a-1,a+r,a+b+1,\dots,a+b+c-1), \ee and
has the extremal cases
\begin{align}
N_{a,b,c}(0)&=M_{a,b,c-1},
&
N_{a,b,c}(b)&=M_{a-1,b,c}.
\end{align}
Yet again, considering ratios we get
\begin{equation}
\frac{N_{a,b,c}(r)}{N_{a,b,c}(0)}=
\binom{a+r-1}{a-1}\binom{b+c-r-1}{c-1}\binom{b+c-1}{c-1}^{-1},
\end{equation}
which in turn leads to
\begin{equation}\label{hexref2}
\frac{N_{a,b,c}(r)}{M_{a,b,c}}=
\binom{a+r-1}{a-1}\binom{b+c-r-1}{c-1}\binom{a+b+c-1}{b}^{-1},
\end{equation}
with $r\in\{0,\dots,b\}$

The Tangent Method is better visualised through the construction of
directed non-intersecting lattice paths, which in this case arise
through a well-known bijection.  For each oblique lozenge, let us draw
a segment connecting the midpoints of its horizontal sides. Clearly,
these segments concatenate to form continuous paths on the triangular
lattice (with steps using only two directions of the lattice, i.e.\ in
fact being directed paths). In the case of the $(n,k)$-trapezoid with
vector $\mathbf{x}$, each lozenge tiling can be now viewed as a
configuration of $k$ non-intersecting paths, connecting the $k$ points
located at $1,\dots,k$ on the short basis with those $k$ points on the
long basis which are at the complement set w.r.t.~$\mathbf{x}$.

\begin{figure}
\begin{center}
\setlength{\unitlength}{17.67pt}
\begin{picture}(14,8)(0.15,0.7)
\put(0.02,0.7){\includegraphics[scale=1.7]{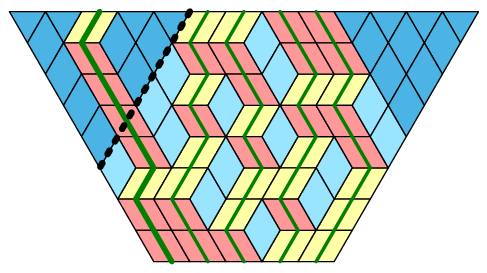}}
\put(2,4){$s$}
\put(2.3,4.1){\line(3,1){.7}}
\put(2.75,8.3){$r$}
\put(6.75,8.3){$b-1$}
\put(2.3,2){$c-1$}
\put(6.7,0.5){$b$}
\put(10.6,2.5){$a$}
\end{picture}
\end{center}
\caption{A lozenge tiling of the $(n,k)$-trapezoid, with
  $(n,k)=(9,6)$.  The locations of the triangular tiles  $\mathbf{x}$ has been
  chosen to provide the refined enumeration $M_{a,b,c}(r)$, with
  $(a,b,c)=(4,5,4)$, and $r=3$. 
  In the non-intersecting lattice path description of lozenge tilings,
  the left-most path, in a thicker green line in the picture, connects
  the left-most horizontal edge on the bottom (short) basis with
  position $r$ on the top (long) basis.  Inside the $(a,b,c)$-hexagon,
  there is a unique oblique lozenge that crosses the dotted line of
  length $a+1$. We denote its distance from the bottom-left side of
  the trapezoid by $s$ (thus $s\in\{0,\dots,a-1\}$). Here $s=1$.}
\label{fig-hex}
\end{figure}
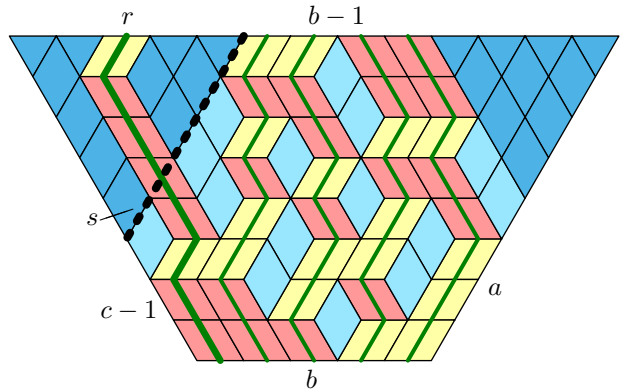

In particular, the choice $\mathbf{x}=(1,\dots,a,a+b+1,\dots,a+b+c)$
gives a lattice path description of the lozenge tilings of the
$(a,b,c)$-hexagon, with $b$ paths connecting the two horizontal sides,
on sequences of consecutive points.  The choice
$\mathbf{x}=(1,\dots,\hat{r},\dots,a+1,a+b+1,\dots,a+b+c)$,
corresponding to our first refinement, $M_{a,b,c}(r)$, see
\eqref{hexref1}, has $b$ paths, starting all contiguous on the short
basis, and arriving all contiguous on the long basis, with the
exception of the left-most path, that arrives at $r$, as illustrated in
Figure~\ref{fig-hex}.

This special path is directed. Thus, in particular, it crosses exactly
once the lattice line that goes in north-east direction, starting from
position $c-1$ on the west side of the trapezoid, this occurring with
an oblique lozenge sheared towards the left, see
Fig.~\ref{fig-hex}. Let us call \emph{cut-line} this special line on
the lattice, and call $s$ the distance of this special lozenge from
the left side of the triangoloid.

Within the ideas of the Tangent Assumption \ref{assum.tangent}, in the
large volume limit this path leaves the Arctic curve tangentially, and
reaches the (given) position $r$ while crossing at a (random) value
$s$, that concentrates on some value $s_{\rm sp}(r)$, up to sub-linear
fluctuations (in fact, of order $N^{\frac{1}{2}}$). As $r$ varies, the
function $s_{\rm sp}(r)$ allows to identify a family of straight
lines, all tangent to the Arctic curve, which thus determine it
through the construction of their caustic.

Any configuration can be decomposed into a part above the cut-line,
and a part below. Say we have $Z_1(s)$ configurations in the part
below, and $Z_2(r,s)$ configurations in the part above. We can
recognise $Z_1(s)$ as the refined enumeration $N_{a,b,c}(s)$, under
the substitution $(a,b,c) \to (c,a,b)$, and, quite trivially, 
$Z_2(r,s)$ just as binomial coefficient:
\begin{align}
Z_1(s)&=N_{c,a,b}(s),
&
Z_2(r,s)&=\binom{a-s}{a-r+1}.
\end{align}
On the other side, the refined enumeration $M_{a,b,c}(r)$, already evaluated
in \eqref{hexref1}, is just
\be
M_{a,b,c}(r)
=
\sum_s
Z_1(s)
Z_2(r,s)
\ee
which leads to the (not completely trivial) identity
\begin{multline}\label{identity_hex}
\sum_{s=0}^{r-1} \binom{a-s}{a-r+1}\binom{c+s-1}{c-1}\binom{a+b-s-1}{b-1}=\\
\binom{a}{r-1}\binom{b+c-1}{c}\binom{a+b+c-1}{a}\binom{a+b+c-r}{c}^{-1},
\end{multline}
holding for any $a$, $b$, $c$ integers, and $r\in\{1,\dots,a+1\}$.

Let us investigate the implications of identity \eqref{identity_hex}
in the `thermodynamic limit', i.e.\ when the hexagon has large size,
and the ratios of the sides is kept fixed.  Let
\begin{align}
a&=\lceil N \alpha\rceil, &
b&=\lceil N \beta\rceil, &
c&=\lceil N \gamma\rceil, &
r&=\lceil N \xi\rceil, &
s&=\lfloor N \eta\rfloor,
\end{align}
with $\alpha,\beta,\gamma,\xi,\eta\in\mathbb{R}$,
$\alpha,\beta,\gamma>0$, and $0<\eta<\xi\leq \alpha$.  When the overall
scale factor $N$ is large, binomials can be replaced with the
dominant term in Stirling formula, while the left-hand side,
interpreted as Riemann sum, can be rewritten as an integral and
evaluated in the saddle-point approximation.  The saddle-point
equation is simply
\begin{equation}
\frac{(\gamma+\eta)(\xi-\eta)}{\eta(\alpha+\beta-\eta)}=1
\end{equation}
with solution 
\begin{equation}\label{hexspesol}
\eta_{sp}=\frac{\gamma\,\xi}{\alpha+\beta+\gamma-\xi},\qquad\xi\in[0,\alpha].
\end{equation}
As $\xi$ varies over $[0,\alpha]$, $\eta_{sp}$ ranges over
$[0,\alpha\gamma/(\beta+\gamma)]$, monotonically. Recalling that $\xi$
and $\eta$ actually parameterize the location of two points in the
`rescaled' plane, the saddle-point solution defines a family of pairs
of points, or equivalently, a family of lines, parameterized by
$\xi\in[0,\alpha]$.  According to the Tangent Assumption, the
corresponding geometric caustic is exactly the Arctic curve we are
looking for (more precisely, its west arc, between the two contact
points with the sides of the hexagon of length $a$ and $c$).

Adapting to the notations of \cite{CLP-98}, we introduce a Cartesian
coordinate system, with origin at the center of the rescaled,
$(\alpha,\beta,\gamma)$-hexagon.  One can check that the sides of the
hexagon lie on the lines
$y=\frac{\sqrt{3}}{2}(2x+\beta+\gamma)$,
$y=\frac{\sqrt{3}}{4}(\alpha+\gamma)$,
$y=\frac{\sqrt{3}}{2}(-2x+\alpha+\beta)$,
$y=\frac{\sqrt{3}}{2}(2x-\beta-\gamma)$,
$y=\frac{\sqrt{3}}{4}(-\alpha-\gamma)$,
$y=\frac{\sqrt{3}}{2}(-2x-\alpha-\beta)$.  In such coordinate system
the two points parameterized by $\xi$ and $\eta$ have coordinates
\begin{equation}
\left(\frac{4\xi-3\alpha-2\beta-\gamma}{4},
\frac{\sqrt{3}}{2}\frac{\alpha+\gamma}{2}\right),
\qquad
\left(\frac{2\eta-\alpha-2\beta-\gamma}{4},
\frac{\sqrt{3}}{2}\frac{2\eta+\gamma-\alpha}{2},
\right),
\end{equation}
respectively. Correspondingly, the family of lines selected by the
saddle-point solution \eqref{hexspesol} has equation
\begin{multline}\label{hexfamily}
2 
\left[(\alpha (\alpha + \beta + \gamma) - (\alpha + \gamma) \xi \right] x 
+
\frac{2}{\sqrt{3}}
\left[\alpha(\alpha+\beta+\gamma) - (3\alpha+2\beta+\gamma) \xi
+2\xi^2\right] y
\\
+
\left[\alpha (\alpha + \beta) (\alpha + \beta + \gamma) - 
2 \alpha (\alpha + \beta + \gamma) \xi + (\alpha + \gamma) \xi^2\right]=0
\end{multline}
with parameter $\xi\in[0,\alpha]$. It is easy to verify that at
$\xi=0$ and $\xi=\alpha$, we recover the lines on which lie the left
sides of the hexagon of rescaled length $\gamma$ and $\alpha$,
respectively.

A tedious but elementary calculation immediately leads to the
parametric form of the corresponding geometric caustic,
\begin{align}
x&=\frac{1}{4} \, \frac{
2 \beta \gamma
+
(\alpha + 2 \beta - \gamma)
(\alpha + \beta - 2 \xi)
+
\left(
1 - \frac{\gamma}{\alpha}
\right)
\left(
1+\frac{\beta}{\alpha + \beta + \gamma}
\right)
\xi^2}
{(\alpha + \beta - 2 \xi) +
\alpha^{-1}
\left(
1-\frac{\beta}{\alpha + \beta + \gamma}
\right)
\xi^2},
\\
y&=\frac{\sqrt{3}}{4}\,
\frac{
2 \alpha \beta
+
(\alpha + \gamma)
(\alpha - \beta - 2 \xi) 
+ 
\left(
1 + \frac{\gamma}{\alpha}
\right)
\left(
1-\frac{\beta}{\alpha + \beta + \gamma}
\right)
\xi^2
}
{(\alpha + \beta- 2  \xi )
+
\alpha^{-1}
\left(
1-\frac{\beta}{\alpha + \beta + \gamma}
\right)
 \xi^2},
\end{align} 
where again $\xi\in[0,\alpha]$. This indeed describes a portion of the
ellipse inscribed in the rescaled hexagon, namely that arc delimited
by the contact points of the ellipse with the two forementioned sides
of the hexagon.  Eliminating the parameter $\xi$, we obtain the
equation $E_{\alpha,\beta,\gamma}(x,y)=0$, where
$E_{\alpha,\beta,\gamma}(x,y)$ is the polynomial
\begin{equation}
3\alpha\beta\gamma(\alpha+\beta+\gamma)
-3(\alpha+\gamma)^2 x^2
+2\sqrt{3}(\alpha+\beta+\gamma)(\alpha-\gamma)xy
-[(\alpha+2\beta+\gamma)^2-4\alpha\gamma]y^2
\end{equation}
introduced in
\cite{CLP-98}, and whose zero-set is indeed the ellipse inscribed in
the $(\alpha,\beta,\gamma)$-hexagon.

Observe that, although in principle the Tangent Method shall have
derived only one portion of the Arctic curve, the polynomial equation
above describes the full Arctic ellipse of the model. As mentioned
above, the possibility of extending one arc to the full curve just by
analytic continuation is a special feature of free-fermionic systems
\cite{KO-05}, which, as seen also from our explicit results on the
six-vertex model, unfortunately seems to fail for other universality
classes.

Let us stress the fact that in \cite{CLP-98} the authors perform the
analysis of the asymptotic behaviour of a much more complex refined
enumeration of the lozenge tilings. On one side, that derivation is
much more complicated than ours. But, most relevant, on the other side
it allows to derive the whole limit shape of the model, where our
restriction of the analysis to a suitably-chosen boundary refinement,
with just one parameter, allows to obtain just the Arctic curve,
i.e.\ the frozen boundary of the limit shape.  It was not clear
\emph{a priori}, from \cite{CLP-98}, that a shorter track existed to
extract the Arctic curve without solving the full limit-shape problem.

\section{Some results for the six-vertex model on a
  triangoloid}
\label{app.triangoloid}

\noindent
The six-vertex model at ice-point, $\fraka=\frakb=\frakc=1$, on the
$n\times n$ square lattice with domain wall boundary conditions can be
reformulated, through a simple bijection, in terms of
\emph{fully-packed loops} (FPL) on the same underlying graph, with
\emph{alternating} boundary conditions.  This relation plays a crucial
role in the formulation \cite{RS-04} and proof \cite{CS-11} of (the
dihedral case of) the Razumov--Stroganov correspondence, which
concerns the enumeration of these configurations according to their
\emph{link pattern}.

A major ingredient in the proof is the notion of \emph{gyration}, an
operation that can be performed on FPL configurations, and was used in
\cite{Wi-00} to prove the dihedral symmetry of FPL on the square
domain.  The proof of the Razumov--Stroganov correspondence given in
\cite{CS-11} for the square domain actually generalizes to a wider
class of domains, called \emph{dihedral domains}, provided that the
gyration operation induces dihedral symmetry \cite{CS-14}. A corollary
of the correspondence is that the enumeration of configurations in a
dihedral domain with perimeter $4n$ factorises into $A_n$, times a
factor, specific to the domain, that counts configurations with a
given `rainbow' link pattern.

The triangoloid is a particular instance of a dihedral domain.
Alternating boundary conditions for FPL on the $n\times n$ lattice
extend naturally to any dihedral domain. In the case of the
triangoloid, they translate for the six-vertex model into the
adaptation of domain wall boundary conditions that is described in
Section \ref{sec.triangoloid.1}.

In this case, the configurations with rainbow link pattern are in
bijection with lozenge tilings of a $(a,b,c)$-hexagon, from which,
calling $A_{a,b,c}$ the number of configurations of the model on the
$(a,b,c)$-triangoloid (that is, the partition function at ice-point),
as described in \cite[Sect.~4.2]{CS-14} (and with a crucial use of 
\cite[Sect.~3]{PZJinffam}), one finds 
\begin{equation}
\label{trienum}
A_{a,b,c}=A_n M_{a,b,c},\qquad n=a+b+c,
\end{equation}
where $A_n$ and $M_{a,b,c}$ are the number of ASM of size $n$, see
\eqref{ASMenum}, and of lozenge tilings of the $(a,b,c)$-hexagon, see
\eqref{macmahon}, respectively.

Yet again, as in Appendices \ref{app.asm} and \ref{app.hex}, we can
consider refined enumerations.
Let $r$ denote the location of the unique thick edge on the bottom row
of the triangoloid, counted from the left, $r\in\{1,\dots,a+2b+c\}$.
Let $A_{a,b,c}(r)$ be the number of six-vertex model configurations on
the $(a,b,c)$-triangoloid, refined according to $r$.

A result of \cite{CS-14}, remarkably related to the fact that a
refined version of the Razumov--Stroganov correspondence holds, is
that
\begin{equation}\label{triref}
A_{a,b,c}(r)=\sum_{s=1}^{r} A_n(s)N_{c,b,a}(r-s),
\qquad r\in\{1,\dots,a+2b+c\},
\end{equation}
where $n=a+b+c$, while $A_n(s)$ and $N_{c,b,a}(r)$ denote the refined
enumerations of ASM of size $n$, see \eqref{ASMref}, and of lozenge
tilings of the $(c,b,a)$-hexagon, see \eqref{hexref2}, respectively.
Note that, even if these two quantities are in principle defined only
for $s\in\{1,\dots,n\}$, and for $r\in\{0,\dots,b\}$, respectively, in
\eqref{triref} we use the convention that the corresponding
expressions \eqref{ASMref} and \eqref{hexref2} just vanish out of the
proper range.

The quantity of interest for the determination of the Arctic curve is
the boundary correlation function $H_{a,b,c}^{(r)}$ defined in
Section~\ref{sec.triangoloid}, and given by
\begin{equation}\label{triHNr}
H_{a,b,c}^{(r)}:=\frac{
A_{a,b,c}(r)}{A_{a,b,c}}.
\end{equation}
A simple calculation leads to the
expression~\eqref{refined-triangoloid}.

\section{Defect lines in the six-vertex model}
\label{app.gaugeZ2}

\noindent
Consider the six-vertex model with spin-reversal symmetry of the
weights (i.e., with weights $\fraka$, $\frakb$ and $\frakc$ instead of
the more general $w_1$, \ldots, $w_6$). In this case we have an
obvious $\mathbb{Z}_2$ symmetry: if we reverse the boundary
conditions, we get the same partition function. As is the case in all
models on planar graphs with such a property, this global symmetry can
be raised to a gauge invariance for the \emph{frustration} of the
associated versions of the model in presence of (anti-ferromagnetic)
defects. The prototype example of this feature is, yet again, the
Ising Model at zero magnetic field, where this has been first
discussed by Toulouse \cite{Toul77, FHS78}.

The \emph{defects}, which in Ising correspond to anti-ferromagnetic
bonds, are here edges enriched with a mid-point, which must have
either two incoming, or two outgoing arrows. As thus, these defects
`live' on the edges of the graph (which, being planar, can be seen as
a 2-dimensional cell complex). Two defects on the same edge clearly
act on vertex configurations as if there is no defect at all.  Now,
consider a graph with all vertices of degree 4, and concentrate on a
given vertex. Some of its incident edges have a defect, some other
don't.  From the spin-reversal symmetry we get that, if we add
a defect to all of the edges incident to this vertex, we get the same
partition function, because of the obvious involution 
$w_1 \leftrightarrow w_2$, $w_3 \leftrightarrow w_4$ and
$w_5 \leftrightarrow w_6$ on the local configuration of the vertex. If
we quotient out this invariance, we see that the information about the
defects is contained in the \emph{faces} of the graph. Each face $f$
has associated a variable
$\nu_f \in \mathbb{Z}_2$, which is the number of defects surrounding
the face, modulo 2. It is easily seen that $\sum_f \nu_f =0$, still
modulo 2. We can draw open paths on the dual graph connecting the
faces with $\nu=1$, in a whatever pairing and through arbitrary
trajectories, and even add closed paths on the dual graph, and then
put a defect on each edge that has been crossed by these lines. The
resulting partition function depends only on $\nu$, and not on the
specific choice of paths.

In presence of one boundary (i.e.\ of vertices of degree 1, all
adjacent to the same face), with fixed boundary conditions, the
external face takes the value of $\nu$ required to have
$\sum_f \nu_f =0$ and then, if this is $1$, we have a defect line
reaching the boundary, at some position between two external edges. If
we move the endpoint of the defect line along the boundary, the
boundary conditions are reversed in the interval along which this
endpoint has been slided. If we perform a full turn of the external
face, thus, the boundary conditions are reversed, and it takes two
turns to go back at the initial data. But this does not cause any
contraddiction, as in fact, as we said above, the partition function
is symmetric under reversal of fixed boundary conditions.

The triangoloid domain discussed in Section \ref{sec.triangoloid} has
a (very moderate but non-zero) presence of such defects.  In that
case, there is exactly one internal face with $\nu_f=1$, namely the
unique triangle, that thus produces one defect line reaching the
boundary. In Section \ref{sec.triangoloid} we made the simplest
possible choice, also in relation to the chosen system of coordinates
(this choice, by the way, breaks a $D_3$ dihedral covariance of the
model under permutations of $(a,b,c)$ parameters, which is manifest in
the gauge-invariant formulation of the defects).  Nonetheless, it is
useful to keep in mind that, in this domain, we have the forementioned
gauge covariance under the deformation of this defect line.

%
%


\end{document}